\documentclass[onecolumn,pre,floats,aps,amsmath,amssymb,nofootinbib]{revtex4-2}
\usepackage{comment}
\usepackage[utf8]{inputenc}
\usepackage{mathtools}
\usepackage{bbm}
\usepackage{cancel} 
\usepackage{dcolumn}% Align table columns on decimal point
\usepackage{blindtext}
\usepackage{xcolor}
\usepackage{amsthm}
\usepackage{bm}% bold math
\usepackage{hyperref}% add hypertext capabilities
\usepackage{graphicx}
\usepackage{graphics,epsfig}
\usepackage{subcaption}
\usepackage{verbatim}
\usepackage{microtype}
\usepackage{adjustbox}
\usepackage{slashed}
\usepackage{physics}
\newcommand{\be}{\begin{equation}}
\newcommand{\ee}{\end{equation}}
\newcommand{\bea}{\begin{eqnarray}}
\newcommand{\eea}{\end{eqnarray}}
\newcommand{\bs}{\boldsymbol}
\newcommand{\beq}{\begin{equation}}
\newcommand{\eeq}{\end{equation}}

\newcommand{\msbar}{\overline{\footnotesize\textrm{MS}}}

\newcommand{\gev}{{\rm GeV}}
\newcommand{\mev}{{\rm MeV}}

\usepackage{simpler-wick}
\usepackage{comment}
\usetikzlibrary{decorations.markings}
\tikzset{W->-/.style={decoration={
  markings,
  mark=at position 0.5*\pgfdecoratedpathlength+2pt with
  {\draw[-latex] (-2pt,0pt) -- (1pt,0pt);}},postaction={decorate}},
  W-<-/.style={decoration={
  markings,
  mark=at position 0.5*\pgfdecoratedpathlength with
  {\draw[latex-] (-2pt,0pt) -- (1pt,0pt);}},postaction={decorate}}
  }
\newif\ifWickBelow
\WickBelowfalse
\pgfkeys{
  /simplerwick/.cd,
  arrows/.store in=\LstWickArrows,
  arrows={-,-,-,-,-,-,-,-,-},
  arrows/.initial={-,-,-,-,-,-,-,-,-}, % the # of contractions is bounded by 9
  below/.code={\WickBelowtrue},
}

\makeatletter
\def\swick@end#1#2{
  \swick@setfalse@#1
  \tikzexternaldisable
  \begin{tikzpicture}[remember picture, baseline=(swick-close#1.base)]
    \node[use as bounding box, inner sep=0pt, outer sep=0pt] (swick-close#1) {$\displaystyle #2$};
  \end{tikzpicture}
  \tikz[remember picture, overlay]
{
\foreach \W@X[count=\W@C] in \LstWickArrows
{\ifnum\W@C=#1
\xdef\myW@style{\W@X}
\fi}
\ifWickBelow
    \draw[\myW@style] ($(swick-open#1.south) + (0, -3pt)$) 
          -- ($(swick-open#1.base) + (0, -\swick@offset) + #1*(0, -\swick@sep)$) 
          -- ($(swick-close#1.base) + (0, -\swick@offset) + #1*(0, -\swick@sep)$) 
          -- ($(swick-close#1.south) + (0, -3pt)$);
\else
    \draw[\myW@style] ($(swick-open#1.north) + (0, 3pt)$) 
          -- ($(swick-open#1.base) + (0, \swick@offset) + #1*(0, \swick@sep)$) 
          -- ($(swick-close#1.base) + (0, \swick@offset) + #1*(0, \swick@sep)$) 
          -- ($(swick-close#1.north) + (0, 3pt)$);
\fi}
  \tikzexternalenable}
\makeatother

\newcommand{\RomatreINFN}{Istituto Nazionale di Fisica Nucleare, Sezione di Roma Tre,\\ Via della Vasca Navale 84, I-00146 Rome, Italy}

\newcommand{\Orsay}{IJCLab, P\^ole Th\'eorie (Bat.~210), CNRS/IN2P3 et Universit\'e,\\ Paris-Saclay, 91405 Orsay, France}
\newcommand{\CyprusI}{Computation-based Science and Technology Research Center, The Cyprus Institute,\\20 Konstantinou Kavafi Street, 2121 Nicosia, Cyprus}

\begin{document}
\title{Lattice QCD determination of the normalization of the 
leading-twist photon distribution amplitude and susceptibility of the quark condensate} 
\author{D.\,Be\v{c}irevi\'c}\affiliation{\Orsay}
\author{G.\,Gagliardi}\affiliation{\RomatreINFN}
\author{F.\,Sanfilippo}\affiliation{\RomatreINFN}
\author{S.\,Bacchio}\affiliation{\CyprusI}

\date{\today}

\begin{abstract}
The normalization of the leading-twist photon distribution amplitude (DA), $f_{\gamma}^{\perp}$, is an important ingredient in the study of exclusive processes involving the photon emission by means of QCD sum-rules. In this paper we determine  the up- , down-  and strange-quark contribution to $f_{\gamma}^{\perp}$ by exploiting its relation to the zero-momentum two-point correlation function of the electromagnetic current $J_{\rm em}^{\mu}$ and the electric component of the tensor current $T^{\mu\nu}$. To that end we employ the gauge ensembles obtained by using $N_{f}=2+1+1$ Wilson-Clover twisted-mass quark flavors, generated by the Extended Twisted Mass (ETM) Collaboration, and after adding all sources of systematic uncertainties, we obtain a total error of $1.5\%$ and $3.5\%$, respectively, for the light- ($u$ and $d$) and strange-quark contribution to $f_{\gamma}^{\perp}(2~{\rm GeV})$ in the $\overline{\mathrm{MS}}$ scheme, thus improving their accuracy by a factor of $2.3$ and $2.8$, respectively. For the strange-quark contribution $f_{\gamma,s}^{\perp}(2~{\rm GeV})$, we observe a discrepancy with respect to previous lattice calculations. By combining our result with the world average lattice value of the chiral condensate, we obtain for the susceptibility of the quark condensate $\chi_d^{\msbar} (2\, \mathrm{GeV}) \simeq \chi_u^{\msbar} (2\, \mathrm{GeV}) =2.17(12)~{\rm GeV^{-2}}$.
\end{abstract}

\maketitle

\section{Introduction}

QCD sum rules (QCDSR) proved to be an efficient method to get the phenomenologically interesting information about the nonperturbative dynamics of hadrons. For the description of electromagnetic properties of hadrons, including the processes with a photon emission, the authors of Refs.~\cite{Balitsky:1983xk,Ioffe:1983ju} modified the standard QCDSR methodology~\cite{Shifman:1978bx} and studied the correlation functions in an external electromagnetic field. From the initial application in computing the nucleon magnetic moments and the radiative $\Delta \to p\gamma$ decay rate, the approach has been broadly employed in numerous applications including the hadronic correction to the $Z^\ast\gamma^\ast \gamma$-vertex, entering the electroweak contribution to $(g-2)_\mu$~\cite{Czarnecki:2002nt}, computation of the neutron electric dipole moment~\cite{Pospelov:2005pr,Narison:2008jp}, studies of chirality violation in the exclusive photoproduction of hard dijets~\cite{Braun:2002en}, evaluation of the hadronic matrix element of the radiative decays of heavy-light mesons~\cite{Colangelo:2005hv,Rohrwild:2007yt} as well as many other processes involving the leading-twist photon distribution amplitude (DA)~\cite{Ball:2002ps,Anikin:2018fom}. A key nonperturbative quantity needed in this description is the susceptibility of the quark condensate, $\chi$, which measures the response of the chiral condensate to the presence of an external electromagnetic field. As we shall see, it is related to the normalization of the leading-twist photon DA, $f^\perp_{\gamma ,q}$, as 
$f^\perp_{\gamma,q} = \chi_q \langle \bar q q\rangle$, where each quantity depends on the renormalization scale in a given renormalization scheme, and $q$ stands for the (light) quark flavor. It appears that it is more convenient to compute $f^\perp_{\gamma,q}$ directly on the lattice and then combine the result with a known value of $\langle \bar q q\rangle^{\msbar}(2\,\mathrm{GeV})$~\cite{FlavourLatticeAveragingGroupFLAG:2021npn} to extract $\chi_q^{\msbar}(2\,\mathrm{GeV})$. The estimate based on the vector meson dominance for the case of lightest flavors in the isospin limit $m_u=m_d\equiv m_\ell$~\footnote{In this approximation the $\rho$-meson pole saturates the sum in the expression given in Ref.~\cite{Ball:2002ps}, so that we have $\chi_\ell =  -f_\rho f_\rho^T/(m_\rho \langle \bar q q\rangle)$. In the numerical evaluation we then use $\langle \bar q q\rangle= - (272(5)~\mev)^3$~\cite{FlavourLatticeAveragingGroupFLAG:2021npn}, and $f_\rho^T/f_\rho = 0.68(2)$~\cite{RBC:2007yjf}, as obtained in lattice QCD, together with $f_\rho^\mathrm{exp}\simeq 208$~MeV and $m_\rho=0.77$~GeV~\cite{ParticleDataGroup:2022pth}.} yields $\chi_\ell^{\msbar}(2\,\mathrm{GeV})=  1.9(1)~\gev^{-2}$, in good agreement with the QCDSR result, $\chi_\ell^{\msbar}(2\,\mathrm{GeV})= 2.1(2)~\gev^{-2}$~\cite{Ball:2002ps}, but both being considerably different from the values advocated in Refs.~\cite{Vainshtein:2002nv,Cata:2009fd}. This situation clearly called for a lattice QCD determination of this quantity. The first such attempts were made in a setup with $N_{f}=2$~\cite{Buividovich:2009ih} and $N_{f}=2+1$~\cite{Braguta:2010ej} but without accounting for the effects of renormalization. Another attempt was made in Ref.~\cite{McNeile:2012xh}, but a detailed lattice QCD study of $f^\perp_{\gamma,q}$ with $N_{f}=2+1$ staggered quark flavors was performed in Ref.~\cite{Bali:2012jv}, and then considerably improved in Ref.~\cite{Bali:2020bcn} where, in the vanishing quark mass limit, the resulting value of $f^\perp_{\gamma,\ell}(2\,\mathrm{GeV})=-(45.4\pm 1.6)$~MeV has been quoted. That value then corresponds to $\chi_\ell^{\msbar}(2\,\mathrm{GeV})= 2.3(1)~\gev^{-2}$. In this paper we calculate $f^\perp_{\gamma,q}$, with $q\in \{u,d,s\}$, from the two-point correlation functions computed on the gauge field configurations generated by using the twisted mass QCD on the lattice with $N_{f}=2+1+1$ Wilson-Clover dynamical quarks~\cite{Frezzotti:2000nk}.

The ensembles of gauge field configurations that we use correspond to three lattice spacings $a\in [0.056,0.08]~{\rm fm}$ and the spatial volumes $V=L^{3}$ with $L \in [5.0, 7.7]~{\rm fm}$~\cite{ExtendedTwistedMass:2021gbo, ExtendedTwistedMass:2021qui, Alexandrou:2022amy,Alexandrou:2018egz}. After taking the continuum and infinite-volume limits, and by adding all sources of systematic uncertainties in quadrature, we obtain a total error of $1.5\%$ for $u,d$ and $3.5\%$ for the $s$ quark. That represents the improvement by a factor of more than $2$ over the previous lattice QCD estimate of these quantities. Our final results in the $\overline{\mathrm{MS}}$ scheme are: 
\begin{align}
f_{\gamma,u}^{\perp}(2~{\rm GeV})   &= -43.73(64)~{\rm MeV}~, \\[8pt]
f_{\gamma,d}^{\perp}(2~{\rm GeV})   &= -44.45(74)~{\rm MeV}~, \\[8pt]
f_{\gamma,s}^{\perp}(2~{\rm GeV})   &= -51.0(1.8)~{\rm MeV}~,
\end{align}
which include the nonperturbative subtraction and multiplicative renormalization constants computed in the RI''-MOM scheme, cf. Ref.~\cite{Alexandrou:2022dtc}, and then converted to the $\msbar$ scheme by using the $4$-loop perturbative expressions available in Ref.~\cite{Gracey:2022vqr}.

\section{Normalization of the photon DA and its relation to the magnetic susceptibility of the quark condensate}

Let us first consider the coupling of a photon to the tensor density $T_{q}^{\alpha\beta}(x,\mu) = Z_{T}(\mu)\,\bar{q}(x)\sigma^{\alpha\beta}q(x)$, and introduce $f_{\gamma,q}^{\perp}(p^{2},\mu)$ as
\begin{align}
\label{eq:def_fgamma}
\langle 0 | T_{q}^{\alpha\beta}(0,\mu) | \gamma(p^{2}, \varepsilon_{\lambda}) \rangle  = ie_{q}\,f_{\gamma, q}^{\perp}(p^{2}, \mu)( \varepsilon^{\alpha}_{\lambda}p^{\beta} - \varepsilon^{\beta}_{\lambda}p^{\alpha})~,
\end{align}
where $\varepsilon_{\lambda}$ is the photon polarization, we use $\sigma^{\mu\nu}= (i/2)[\gamma^{\mu}, \gamma^{\nu}]$, and $e_{q}= e Q_q$ is the electric charge of the $q-$quark, with $e=\sqrt{4\pi \alpha_\mathrm{em}}$, and $Q_u=+2/3$, $Q_{d,s}=-1/3$. Note that $Z_{T}(\mu)$ is the scale- and scheme-dependent renormalization constant of the tensor density. For a real photon $p^{2}=0$, $f_{\gamma,q}^{\perp}(0, \mu)\equiv f_{\gamma,q}^{\perp}(\mu)$ corresponds to the normalization  of the $q$-quark contribution to the leading-twist photon DA~\cite{Ball:2002ps}.~\footnote{To be more specific, $f_{\gamma,q}^{\perp}(\mu)$ is a factor multiplying the leading twist photon DA defined through the matrix element of a nonlocal operator (near light cone, $x^2=0$), namely 
\begin{equation}
    \langle 0 \vert \bar q(x) \sigma^{\alpha\beta} W_{-x}^x q(-x) \vert \gamma(p^2,\varepsilon_\lambda)\rangle = i e_q \chi \langle \bar q q\rangle (p^\beta \varepsilon^\alpha_\lambda - p^\alpha \varepsilon^\beta_\lambda) \int_0^1 
du e^{i p x (2u - 1)} \phi_\gamma (u),\nonumber
\end{equation}
where $W_x^y$ is the Wilson line. In this way, the distribution is normalized as usual, $\int_0^1 du \phi_\gamma (u) =1$, and 
$f_{\gamma,q}^{\perp}(\mu)= \chi(\mu)\, \langle \bar q q\rangle(\mu)$.} 

To leading order in the electromagnetic interaction, the matrix element defined in Eq.~(\ref{eq:def_fgamma}) can be expanded as~\footnote{Our choice of the sign is consistent with
$D_{\mu} = \partial_{\mu} - ie_{q}A_{\mu}$, which is the same convention used in Ref.~\cite{Ball:2002ps} for the covariant derivative.}
\begin{align}
\langle 0 | T_{q}^{\alpha\beta}(0,\mu) | \gamma(p^{2}, \varepsilon_{\lambda})\rangle_{\rm QCD+QED}  &= i\int d^{4}x~\langle 0 | {\rm T}\left\{ T_{q}^{\alpha\beta}(0,\mu) J_{\rm em}^{\nu}(x) A_{\nu}(x) \right\} | \gamma(p^{2}, \varepsilon_{\lambda}) \rangle + \mathcal{O}(\alpha_{{\rm em}}^{3/2}) \nonumber\\[8pt]
&= \varepsilon_{\lambda, \nu}H_{q}^{\alpha\beta\nu}(p)  + \mathcal{O}(\alpha_{{\rm em}}^{3/2})~.
\end{align}
where $ J_{\rm em}^{\mu} \equiv \sum_{q} e_{q}\bar{q}\gamma^{\mu}q$ is the electromagnetic current, and we introduced the hadronic tensor
\begin{align}
H_{q}^{\alpha\beta\nu}(p) \equiv i\int d^{4}x\,e^{-ipx} \langle 0 | {\rm T}\left\{ T_{q}^{\alpha\beta}(0,\mu)\, J_{\rm em}^{\nu}(x)\right\} | 0 \rangle~.
\end{align}
By virtue of Lorentz invariance the hadronic tensor $H_{q}^{\alpha\beta\nu}(p)$ can be written in terms of a single scalar form factor $H_{q}(p^{2},\mu)$ as 
\begin{align}
H_{q}^{\alpha\beta\nu}(p) \equiv i\int d^{4}x\,e^{-ipx} \langle 0 | {\rm T}\left\{ T_{q}^{\alpha\beta}(0,\mu)\, J_{\rm em}^{\nu}(x)\right\} | 0 \rangle \equiv ie_{q}H_{q}(p^{2},\mu)\left( g^{\alpha\nu}p^{\beta}-g^{\beta\nu}p^{\alpha}\right)~,  
\end{align}
so that, to leading order in the electromagnetic interaction, one arrives at $H_{q}(p^{2},\mu) = f_{\gamma,q}^{\perp}(p^{2},\mu)$.

The form factor $H_{q}(0, \mu)$ is also related to the zero-temperature magnetic susceptibility $\chi_{q}(\mu)$ of the quark condensate $\langle\bar{q}q\rangle(\mu)$. The latter is defined through the relation~\cite{Ball:2002ps}
\begin{align}
\label{eq:def_chi}
\langle 0 | T_{q}^{\alpha\beta}(0,\mu) | 0\rangle_{B} \equiv F^{\alpha\beta}e_{q}\chi_{q}(\mu)\langle \bar{q}q\rangle (\mu)~,
\end{align}
where $F^{\alpha\beta} \equiv \partial^{\alpha}A^{\beta} - \partial^{\beta}A^{\alpha} $
is the electromagnetic field-strength tensor, corresponding to a constant weak background magnetic field $B$, and $\langle \dots \rangle_{B}$ means the expectation value evaluated in such a magnetic background field. 

The connection between $H_{q}(0,\mu)$ and $\chi_{q}(\mu)$ can be easily worked out since
\begin{align}
\langle 0 | T_{q}^{\alpha\beta}(0,\mu) | 0\rangle_{B} = i\int d^{4}x \langle 0 | {\rm T}\left\{T_{q}^{\alpha\beta}(0,\mu)\, J_{\rm em}^{\nu}(x)\right\} | 0\rangle A_{\nu}(x) + \mathcal{O}(\alpha_{\rm em}^{3/2})~.
\end{align}
Choosing  $A_{0}=0, A_{j} = \varepsilon_{jk\ell}x^{k}B^{\ell}/2$, where $\vec{B}$ is the constant and uniform magnetic field, alows us to write, to leading order in the magnetic field,
\begin{align}
\langle 0 | T_{q}^{\alpha\beta}(\mu) | 0\rangle_{B} &= -\varepsilon_{jk\ell} \frac{B^{\ell}}{2}\frac{\partial}{\partial p_{k}}\int d^{4}x~e^{-ipx} \langle 0 | {\rm T}\left\{T_{q}^{\alpha\beta}(0,\mu)\, J_{\rm em}^{j}(x)\right\} | 0\rangle \bigg|_{p^2=0} \nonumber \\[8pt]
&= -e_{q}H_{q}(0,\mu)\,\varepsilon_{jk\ell} \frac{B^{\ell}}{2} \left[  g^{\alpha j}g^{\beta k}- g^{\beta j}g^{\alpha k} \right] = e_{q}H_{q}(0,\mu) F^{\alpha\beta} ~,
\end{align}
which, after a comparison with Eq.~(\ref{eq:def_chi}), gives $\chi_{q}(\mu)\langle \bar{q}q\rangle(\mu)= H_{q}(0,\mu) = f^{\perp}_{\gamma,q}(\mu)$. Similar expressions, relating $\chi_{q}(\mu)$ to current-current correlators have already been derived in Ref.~\cite{Bali:2020bcn}, where the authors considered a rotating magnetic background field and the correlation function between the electromagnetic current ($J_{\rm em}$) and the magnetic part, $T_{q}^{ij}$ ($i\ne j$), of the tensor current.

\subsection{Extracting $H_{q}(0,\mu)$ from Euclidean lattice correlators}

To compute the form factor $H_{q}(0,\mu)$ we will be working in the so-called time-momentum representation and set $p=(p_{0}, 0, 0, 0)$. We can write 
\begin{align}
\frac{1}{3}\sum_{j=1}^3 H_{q}^{0j j}(p) = i\int_{-\infty}^{\infty} dt~e^{-ip_{0}t}\int d^{3}x~\frac{1}{3}\sum_{j}\, \langle 0 | {\rm T}\left\{T_{q}^{0j}(0,\mu)\, J_{\rm em}^{j}(\vec{x},t)\right\} | 0\rangle 
= ie_{q}H_{q}(p_{0}^{2},\mu)\, p_{0}~.
\end{align}
The crucial ingredient that makes this calculation possibile is the observation that for small enough values of $|p_{0}|$ the previous expression can be analytically continued to Euclidean time without encountering any obstruction since QCD has a mass gap.~\footnote{ The analytic continuation is possible for $|p_{0}| < E_{0}$, where $E_{0}$ is the energy of the lightest hadronic state propagating between the two currents. In QCD the threshold energy is $E_{0} = 2M_{\pi}$.} For small $|p_{0}|$ we can thus write:
\begin{align}
\label{eq:an_cont_1}
-\frac{i}{3}\sum_{j}\, H_{q}^{0jj}(p) &= e_{q}H_{q}(p_{0}^{2},\mu) p_{0} = -\int_{-\infty}^{\infty} dt\,e^{-p_{0}t}\,C_{q}(t,\mu)~,
\end{align}
where $C_{q}(t,\mu)$ is the Euclidean lattice correlator
\begin{align}
\label{eq:an_cont_2}
C_{q}(t,\mu) &\equiv \frac{i}{3}\int d^{3}x \,\sum_{j}\,\langle 0 | {\rm T}\left\{ T_{q}^{0j}(0,\mu)\, J_{\rm em}^{j}(\vec{x},t)\right\} | 0 \rangle_{E}~,
\end{align}
and the subscript $E$ indicates that the correlation function is evaluated in the Euclidean theory. The zero-momentum correlator, $C_{q}(t,\mu)$, is the nonperturbative object that we compute by means of QCD simulations on the lattice. The only renormalization constant needed in this work is $Z_{T}(\mu)$ to which we will come back later on. In the following, for notational simplicity, we will drop out the renormalization scale dependence.

After taking the $p_{0}$-derivative of both sides of Eq.~(\ref{eq:an_cont_1}) at $p_{0}=0$ we obtain the key expression:
\begin{align}
\label{eq:kernel}
2\int_{0}^{\infty}dt\, t\, C_{q}(t) = e_{q}\, H_{q}(0)~,
\end{align}
where we used the fact that $C_{q}(t)$ is odd under the Euclidean time reflection.

Notice, however, that the expression~(\ref{eq:kernel}) is well defined only in the chiral limit. For a non-zero quark mass, instead, the contact term between the tensor and electromagnetic currents generates a logarithmic divergence in Eq.~(\ref{eq:kernel}) which must be properly subtracted away. The logarithmic divergence is present already in the free theory where the corresponding correlator $C_q^{\rm free}(t)$, owing to the asymptotic freedom, is a good approximation of the full QCD correlator $C_{q}(t)$ only for very short time separations $t\ll\Lambda_{QCD}^{-1}, m_{q}^{-1}$. The free theory calculation is discussed in detail in Appendix~A. Here we only quote the final result, namely,
\begin{align}
C_q^{{\rm free}}(t) = -{\rm sgn}(t)\,\frac{e_{q}N_{c}}{2\pi^{2}}\, \frac{m^{2}_{q}}{|t|}\, K_{1}(2m_{q}|t|) \underset{0< t\ll m_{q}^{-1}}{\simeq} -\frac{e_{q}N_{c}}{4\pi^{2}}\,\frac{m_{q}}{t^{2}}~, 
\end{align}
where $K_{1}(x)$ is the modified Bessel function of the second kind. Since $C_q^{\rm free}(t) \propto m_{q}/t^{2}$ at small times, the regularized time-integral develops a logarithmic divergence:
\begin{align}
\label{eq:log_divergence}
2\int_{\Lambda^{-1}}^{\infty}dt\, t\, C_q^\mathrm{free}(t) \simeq e_{q}\,c_{0}\,m_{q}\log(m_{q}/\Lambda) + \text{regular terms}~,
\end{align}
with $c_{0} = 2N_{c}/(2\pi)^{2}\simeq 0.15$ for $N_{c}=3$. In the lattice regularization, the ultraviolet cut-off $\Lambda^{-1}$ is given by the lattice spacing $a$. In order to define the form factor $H_{q}(0)$ at non-zero quark mass, it is thus necessary to provide a prescription on how the logarithmic divergence is subtracted away. To this end we adopt a convenient prescription, described in Ref.~\cite{Bali:2012jv}, and define the subtracted (finite) form factor $\overline{H}_{q}$ via
\begin{align}
\overline{H}_{q}(0) = \left( 1 - m_{q}\frac{\partial}{\partial m_{q}}\right)\,H_{q}(0)~.
\end{align}
Note that this prescription leaves the chiral limit unchanged, i.e.,
\begin{align}
\label{eq:subtraction_procedure}
2\int_{0}^{\infty} dt\, t\, \left[ C_{q}(t) - m_{q}\frac{\partial}{\partial m_{q}}C_{q}(t) \right] = e_{q}\overline{H}_{q}(0)~.
\end{align}
As discussed in Refs.~\cite{Bali:2012jv,Bali:2020bcn}, the log-derivative will not only cancel the divergent contribution proportional to $m_{q}\log(am_{q})$, but also the finite terms, proportional to powers of the quark mass $m_{q}$. 
The main result of this paper, which we are going to discuss in the next Sections, is the accurate determination of $\overline{H}_{q}(0)$ for the physical light ($q=u,d$) and strange  ($q=s$) quarks, by making use of the gauge ensembles recently generated by the Extended Twisted Mass (ETM) Collaboration~\cite{ExtendedTwistedMass:2021gbo, ExtendedTwistedMass:2021qui, Alexandrou:2022amy,Alexandrou:2018egz}.

\section{Details of the computation}
Our results have been obtained by using the gauge field configurations generated by the ETM Collaboration employing the Iwasaki gluon action~\cite{Iwasaki:1985we} and $N_{f}=2+1+1$ flavors of Wilson-Clover twisted-mass fermions at maximal twist~\cite{Frezzotti:2000nk}. This framework guarantees the automatic $\mathcal{O}(a)$ improvement of on-shell parity-even quantities~\cite{Frezzotti:2003ni,Frezzotti:2004wz}. Basic information regarding 
the ensembles of gauge field configurations used in this work is collected in Table~\ref{tab:simudetails}. 
For more details the reader is referred to Refs.~\cite{ExtendedTwistedMass:2021gbo,ExtendedTwistedMass:2021qui,Alexandrou:2022amy,Alexandrou:2018egz}. In Table\,\ref{tab:renormalization} we give the values of the renormalization constants needed for the present work. The ensembles listed in Table~\ref{tab:simudetails} correspond to three values of the lattice spacing $a\in [0.056,0.08]\,{\rm fm}$, and the size of the lattice box $L\in [5.09, 7.64]\,{\rm fm}$. The mass of the light sea-quarks has been tuned so as to give (almost) the physical pion mass. For all the ensembles, the strange and the charm sea-quarks are set to within about 5\% of their physical masses, defined through the requirement that, cf. Refs.~\cite{Alexandrou:2018egz,ExtendedTwistedMass:2021qui, Alexandrou:2022amy}, 
\begin{align}
\frac{M_{D_{s}}}{f_{D_{s}}} = 7.9\pm0.1\, , \qquad 
\frac{m_{c}}{m_{s}} = 11.8\pm0.2\,.
\end{align}

\begin{table}
\renewcommand{\arraystretch}{2.2}
\begin{center}
    \begin{tabular}{|c||c|c|c|c|c||c|}
    \hline
    ~ Ensemble ~ & ~~~ $\beta$ ~~~ & ~~~ $(L^3\times T)/a^{4}$ ~~~ & ~~~ $a$\,[fm] ~~~ & ~~~ $am_{\ell}$ ~~~ & ~ $M_{\pi}$\,[MeV] ~ & ~ $L$ [fm] ~ \\
  \hline
  
  B64 & $1.778$ & $64^{3}\times 128$ & $0.07957~(13)$ & $0.00072$ & $140.2~(0.2)$ & $5.09$  \\

 B96 & $1.778$ & $96^{3}\times 192$ & $0.07957~(13)$ & $0.00072$ & $140.2~(0.2)$ & $7.64$ \\ 
  
  C80 & $1.836$ & $80^{3}\times 160$ & $0.06821~(13)$ & $0.00060$ & $136.7~(0.2)$ & $5.46$  \\
  
  D96 & $1.900$ & $96^{3}\times 192$ & $0.05692~(12)$ & $0.00054$ & $140.8~(0.2)$ & $5.46$  \\
  \hline
    \end{tabular}
\end{center}
\caption{\it \small Parameters of the ETM ensembles used in this work: the light-quark bare mass, ($am_\ell = a m_u = a m_d$) and the corresponding pion mass ($M_\pi$), the lattice spacing ($a$), and the spatial size of lattice box $L$. The lattice spacing values are determined in a way  explained in Appendix~B of Ref.~\cite{Alexandrou:2022amy}, i.e., by using $f_\pi^\mathrm{isoQCD} = 130.4(2)~{\rm MeV}$ for the pion decay constant. We refer to~\cite{Alexandrou:2022amy} for more  information regarding the statistics of the used ensembles.}
\label{tab:simudetails}
\end{table}

\begin{center}
\renewcommand{\arraystretch}{2.}
\begin{table}[]
    \centering
    \begin{tabular}{|c||c|c|c|}
    \hline
    ensemble & $Z_{V}$ & $Z_{A}$ & $Z_{T}^{\msbar} (2~{\rm GeV})$ \\
    \hline
    ~ B64 ~ & ~ $0.706379(24)$ ~ & ~ $0.74294(24)$ ~ & ~ $0.847(1)(1)$  \\
    \hline
    ~ B96 ~ & ~ $0.706405(17)$ ~ & ~ $0.74267(17)$ ~ & ~ $0.847(1)(1)$  \\
    \hline
    ~ C80 ~ & ~ $0.725404(19)$ ~ & ~ $0.75830(16)$ ~ & ~ $0.863(1)(2)$ \\
    \hline
    ~ D96 ~ & ~ $0.744108(12)$ ~ & ~ $0.77395(12)$ ~ & ~ $0.887(1)(2)$ \\
    \hline
    \end{tabular}
    \caption{\it \small Renormalization constants (RCs) of the quark bilinear operators corresponding to the ETM ensembles of Table~\ref{tab:simudetails}. $Z_{V}$, $Z_{A}$, and $Z_{T}$ stand for the RCs of the vector, axial and tensor operator, respectively. $Z_{T}$ has been determined nonperturbatively in Ref.~\cite{Alexandrou:2022dtc} in the $RI''$-scheme and then converted to the $\msbar$. The scale independent renormalization constants $Z_{V}$ and $Z_{A}$ have been determined in Ref.~\cite{Alexandrou:2022amy} using the Ward identity method.}
    \label{tab:renormalization} 
\end{table}
\end{center}
We work in a mixed-action framework in which the valence strange quark is discretized as an Osterwalder-Seiler fermion~\cite{Osterwalder:1977pc, Frezzotti:2004wz}. On each ensemble we evaluate the lattice correlator $C_{s}(t)$ for two different values of the valence strange quark mass. The two simulated values, given in Table~\ref{tab:masses}, then allow us to interpolate to the physical strange quark mass which, as in  Refs.~\cite{Alexandrou:2022amy, Frezzotti:2023ygt}, we define by requiring $M_{\eta_{ss'}} = M_{\eta_{ss'}}^{\mathrm{phys}} = 689.9(5)\,{\rm MeV}$~\cite{Borsanyi:2020mff}, 
with $\eta_{ss'}$ being a fictitious pseudoscalar meson made of two mass degenerate quarks $s$ and $s'$ having mass equal to that of the strange quark. Besides the two valence strange quark masses, we employ a second  value of the valence bare light quark mass, indicated in Table~\ref{tab:masses} as $am_{\ell}'$,  different from the sea light quark mass already given in Table~\ref{tab:simudetails}. For each flavor $q$, having results for (at least) two different valence quark masses is necessary in order to compute the derivative $\partial C_{q}(t)/\partial m_{q}$ entering the definition of $\overline{H}_{q}(0)$ in Eq.~(\ref{eq:subtraction_procedure}), as we describe in the next Section.

\begin{figure}
    \centering
    \includegraphics[scale=0.8]{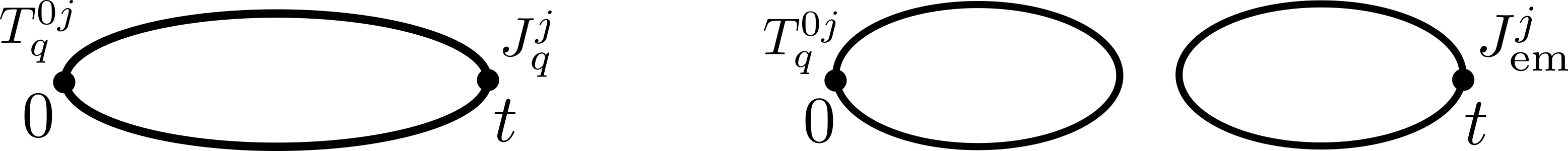}
    \caption{\small\it Graphical representation of the quark-line connected (left diagram) and disconnected (right diagram) contributions to the Euclidean correlator $C_{q}(t)$. \label{fig:Feyn} }
    
\end{figure}

The Euclidean correlators $C_{q=u,d,s}(t)$ in Eq.~(\ref{eq:an_cont_2}) have been evaluated for the aforementioned valence quark masses for each of the ensembles listed in Table~\ref{tab:simudetails}.
Note that $C_{q}(t)$ contains both a connected [$C_{q}^{\rm conn}(t)$] and a disconnected [$C_{q}^{\rm disc}(t)$] contribution. If we define the electromagnetic current as,
\begin{align}
J_{\rm em}^{\mu}(x) = \sum_{q'=u,d,s,c}  J_{q'}^{\mu}(x) , \qquad\qquad J_{q'}^{\mu}(x) = e_{q'} \bar{q}'(x)\gamma^{\mu}q'(x)~,
\end{align}
we can write $C_{q}(t) = C_{q}^\mathrm{conn}(t) + C_{q}^\mathrm{disc}(t)$, with
\begin{align}
C_{q}^{\rm conn}(t) &=  \frac{i}{3}\int d^{3}x \,\langle 0 | {\rm T}\left\{ \wick[arrows={W->-,W-<-},below]{ \c1 T_{q}^{0j}\c2(0,\mu)~\c1 J_{q}^{j}\c2 (\vec{x},t)}\right\} | 0 \rangle_{E}\nonumber \\[8pt]
&=  \frac{e_{q}}{3}\int d^{3}x \, \langle \Tr\left[ \gamma^{0}\gamma^{j}S_{q}(0, (\vec{x},t)) \gamma^{j} S_{q}((\vec{x},t), 0)\right]\rangle_{U}~ \\[10 pt]
C_{q}^{\rm disc}(t) &= \frac{i}{3}\int d^{3}x \,\langle 0 | {\rm T}\left\{ \wick[arrows={W->-,W->-},below]{ \c1 T_{q}^{0j}\c1(0,\mu)~\c2 J_{\rm em}^{j}\c2 (\vec{x},t)}\right\} | 0 \rangle_{E} \nonumber \\[8pt]
&= -\sum_{q'=u,d,s,c} \frac{e_{q'}}{3} \int d^{3}x\, \langle \Tr\left[ \gamma^{0}\gamma^{j}S_{q}(0,0) \right] \Tr \left[ \gamma^{j} S_{q'}((\vec{x},t), (\vec{x},t))\right]\rangle_{U}~,
%-\frac{i}{3}\int d^{3}x \,\langle 0 | {\rm %T}\left\{ \wick[arrows={W->-,W-<-},below]{ %\c1 T_{q}^{0j}\c2(0,\mu)~\c1 J_{\rm %em}^{j}\c2 (\vec{x},t)}\right\} | 0 %\rangle_{E} =
\end{align}
where by $S_{q}(x,y)$ we denote the propagator of the $q$-quark, and $\langle \dots \rangle_{U}$ indicates the average over the $\textrm{SU}(3)$ gauge field configurations, and the trace $\Tr[\dots ]$ is taken over the color and spin indices. The connected and disconnected contributions are sketched in Figure~\ref{fig:Feyn}.
While the connected part $C_{q}^{\rm conn}(t)$ receives the contribution from the $q-$quark component of the electromagnetic current, the disconnected diagram receives contributions from all quark flavors. The latter vanishes exactly in the $\rm{U}(3)$ limit $m_{u}=m_{d}=m_{s}$ of the three-flavor theory, 
due to $e_{u}+e_{d}+e_{s}=0$. In absence of the disconnected contribution, the form factors $\bar{H}_{u}(0)$ and $\bar{H}_{d}(0)$ corresponding to the up- and down-quark are exactly equal. Indeed, for $m_{u}=m_{d}\equiv m_{\ell}$, one has $C_{u}^{\rm conn}(t)/e_{u} = C_{d}^{\rm conn}(t)/e_{d}$ (see also Eq.~(\ref{eq:kernel})). The disconnected part of the correlator, for $m_{u}=m_{d}\equiv m_{\ell}$, satisfies instead $C_{u}^{\rm disc}(t)= C_{d}^{\rm disc}(t)$, which produces a shift of the form factor of the down quark which is two times larger (and opposite in sign) with respect to that of the up quark. 

\begin{figure}
\includegraphics[scale=0.35]{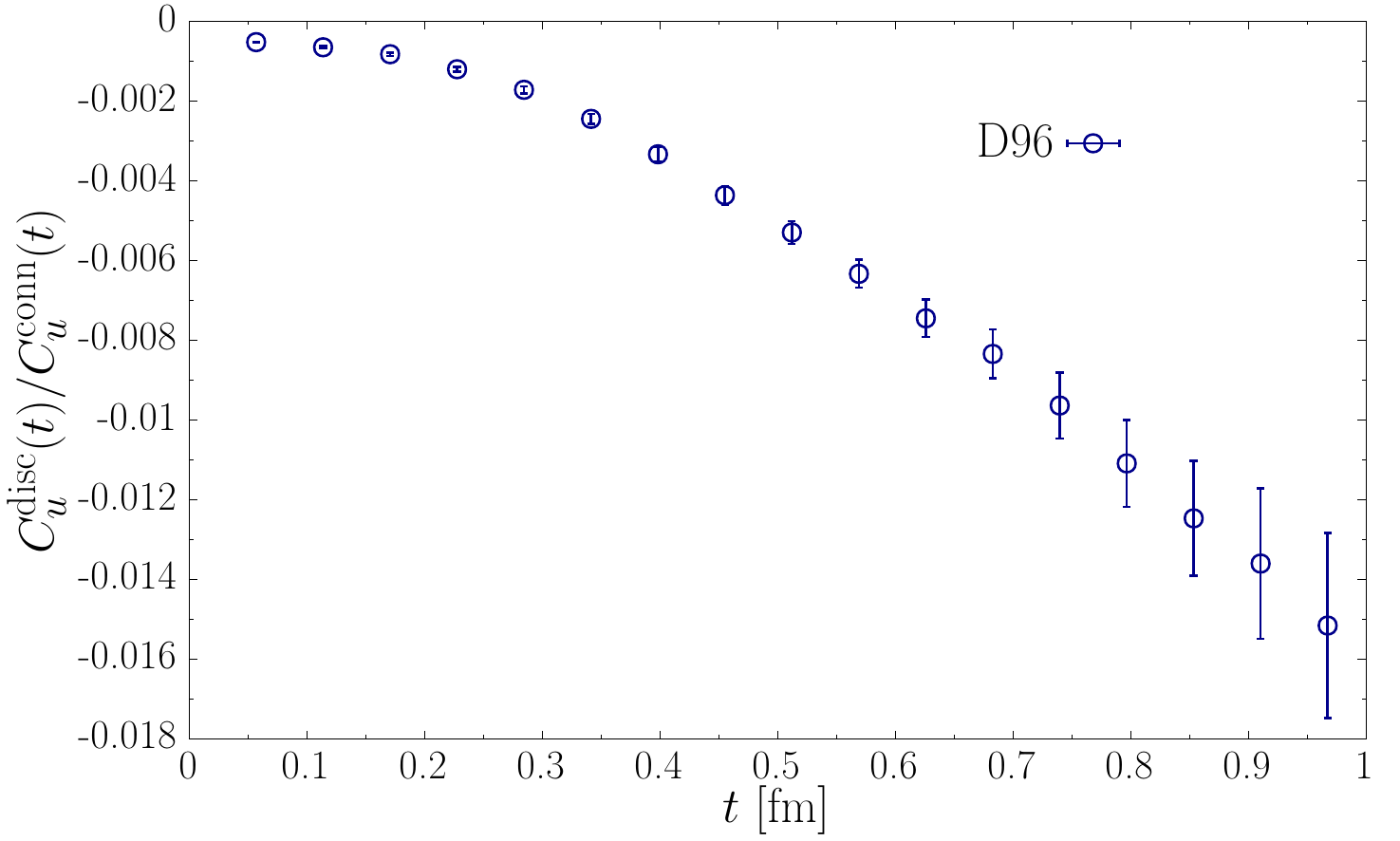}
\includegraphics[scale=0.35]{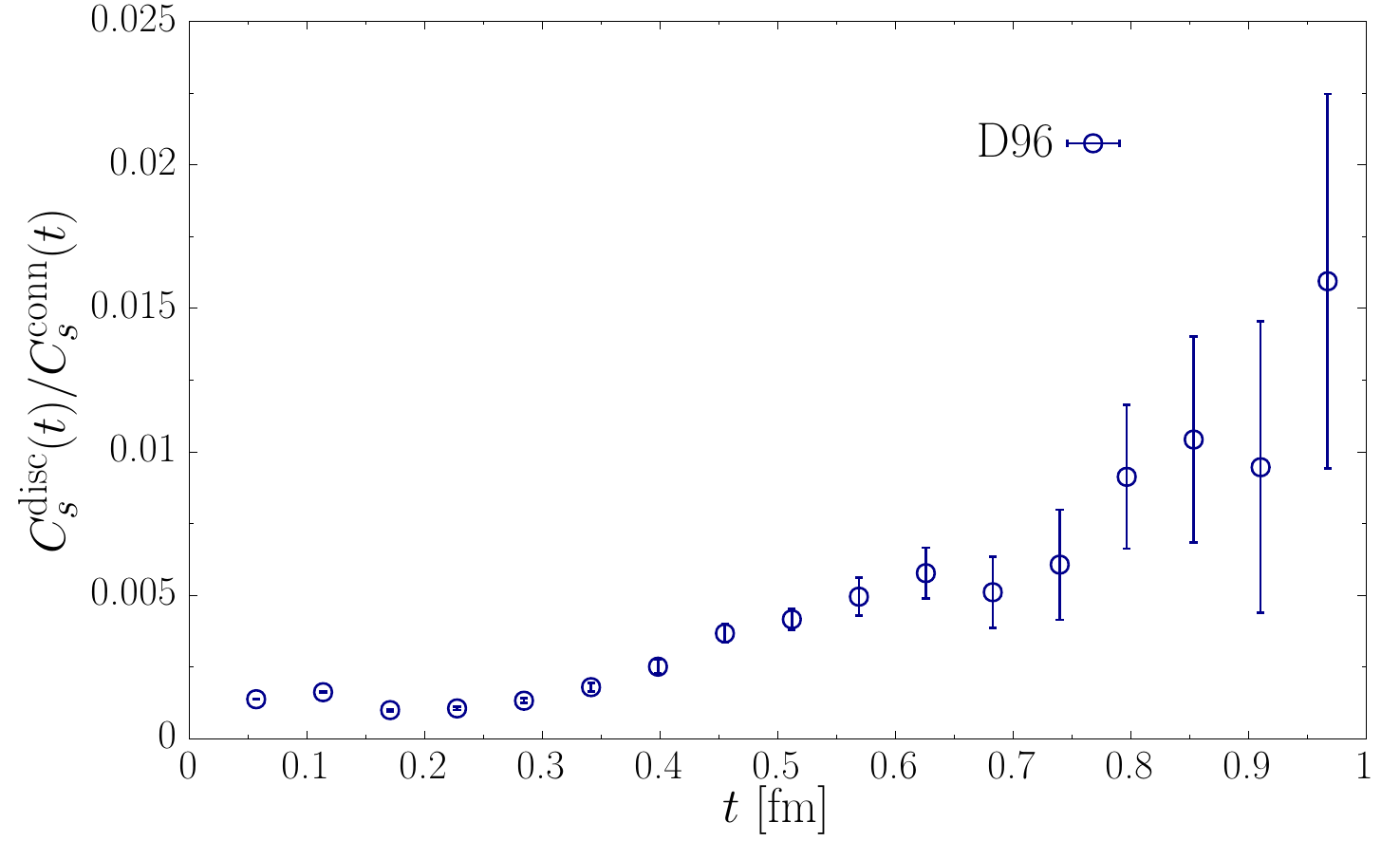}
\caption{\small\it Ratio between the disconnected ($C_{q}^{\rm disc}(t)$) and connected ($C_{q}^{\rm conn}(t)$) contributions to the correlator $C_{q}(t)$ for $q=u$ (left-plot) and $q=s$ (right-plot). The data correspond to the results obtained on the D96 ensemble.   \label{fig:conn_disc}  }
\end{figure}

For the connected part, the inversions of the Dirac operator have been performed using $\mathcal{O}(10^{3})$ and  $\mathcal{O}(60)$ stochastic sources, respectively for the light and the strange quarks. The sources are randomly distributed over time, diagonal in spin and dense in color. Various noise-reduction techniques are instead employed for disconnected loops. These are the one-end-trick~\cite{McNeile:2006bz}, the exact deflation of low-modes~\cite{Gambhir:2016uwp}, and hierarchical probing~\cite{Stathopoulos:2013aci}. In contrast to the connected contribution, the disconnected loops have been computed using a single value of the valence quark mass for each of the $N_{f}=2+1+1$ active quark flavors. For the light quarks $u,d$,  this mass is equal to the sea quark mass $am_{\ell}$, cf. Table~\ref{tab:simudetails}, while for the strange and charm quarks the simulated valence quark masses are obtained by tuning the $\Omega$ and $\Lambda_c$ baryons to their physical values respectively. In Figure~\ref{fig:conn_disc} we show the ratio between the disconnected ($C_{q}^{\rm disc}(t)$) and connected ($C_{q}^{\rm conn}(t)$) contributions to the correlator $C_{q}(t)$, for $q=u,s$. Clearly the disconnected contribution is orders of magnitude smaller than the connected one. After inspection, we find that the disconnected term produces a shift of the form factor $H_{q}(0)$ of order $0.5\%, 1\%$ and $0.3\%$, respectively for $q=u,d$, and $s$. 

Finally, another feature of our calculation is that the (completely dominant) connected part of the  light- and strange-quark correlators are always evaluated employing two distinct discretized versions of the local tensor and vector currents given by,
\begin{align}
T_{q}^{0j, {\rm tm}} &= Z_{T}(\mu)\bar{q}_{+}\gamma^{0}\gamma^{j}q_{-}\,,\qquad\qquad J_{q}^{j, {\rm tm}} = Z_{A}\bar{q}_{+}\gamma^{j}q_{-}\,, \\[8pt]
T_{q}^{0j, {\rm OS}} &= Z_{T}(\mu)\bar{q}_{+}\gamma^{0}\gamma^{j}q_{+}\,,\qquad\qquad J_{q}^{j, {\rm OS}} = Z_{V}\bar{q}_{+}\gamma^{j}q_{+}\,,
\end{align}
where $Z_{A}$ and $Z_{V}$ are the (finite) renormalization constants of the axial and vector currents, which in twisted-mass QCD are chirally rotated with respect to (w.r.t.) the ones of the standard Wilson fermions. In the previous equation, the subscript $\pm$ on the quark-fields, corresponds to a specific choice of the Wilson $r-$parameter (see e.g.~\cite{Alexandrou:2022amy} and references therein), given by 
\begin{align}
r_{q_{\pm}} = (-1)^{\pm}~.
\end{align}
Since the twisted Wilson term, accompanied by the appropriate critical mass counterterm (for both sea
and various valence quark fields), is a dimension-five irrelevant operator, the currents $T_{q}^{{\rm tm}}, J_{q}^{{\rm tm}}$ and $T_{q}^{{\rm OS}}, J_{q}^{{\rm OS}}$, when plugged into the expression~(\ref{eq:an_cont_2}), produce Euclidean correlators having the same continuum limit. Notice also that the label ``tm'' stands for twisted-mass and ``OS'' for Osterwalder-Seiler regularization.  
At finite lattice spacing, however, the two correlators are affected by different discretization effects, allowing one to approach the continuum limit from two different directions. The results shown in Figure~\ref{fig:conn_disc} have been obtained by using $C_{q}^{\rm conn}$ in the ``tm'' regularization. 
In the following, we will present the results obtained using both the ``tm'' and the ``OS'' currents, which, as it will be discussed, can simultaneously be used to reach the continuum limit and achieve a better control of extrapolation. Moreover, we will oftentimes adopt the notation $X^{\rm tm\, (OS)}$, to indicate that a given lattice observable $X$ has been evaluated using the ${\rm tm}\,({\rm OS})$ current. We emphasize once again that the  values of $Z_{A}$ and $Z_{V}$ for each of our ensembles listed in  Table~\ref{tab:simudetails} are provided in Table~\ref{tab:renormalization}.

\subsection{Evaluating the derivative $\partial C_{q}(t)/ \partial m_{q}$ \label{sec:derivative}}

We now discuss a delicate point regarding the derivative $\partial C_{q}(t)/\partial m_{q}$ for $q=u, d, s$. We start by splitting it to a valence- and a sea-quark contribution as 
\begin{align}
\label{eq:der_contribs}
\frac{\partial C_{q}(t)}{\partial m_{q}} = \frac{\partial C_{q}(t)}{\partial m_{q}}\bigg|_{\rm val} + \frac{\partial C_{q}(t)}{\partial m_{q}}\bigg|_{\rm sea}~.  
\end{align}
In order to properly define the two contributions, we start from the path integral representation of a generic observable $\mathcal{O}(m_{q})$. For the sake of simplicity and to avoid unnecessary complications, we consider the continuum action of a single quark field $q(x)$ of the bare quark mass, $m_{q}$, and we avoid explicitly writing the gauge field. We have
\begin{align}
\langle \mathcal{O}(m_{q}) \rangle = \frac{1}{\mathcal{Z}(m_{q})}\int [d\bar{q}dq] \exp\left\{-S_{0}[\bar{q},q]  -m_{q}\int d^{4}x \, \bar{q}q(x)\right\} \mathcal{O}(\bar{q},q, m_{q})~,
\end{align}
where $S_{0}$ is the action in the massless limit and $\mathcal{Z}(m_{q})$ is the partition function of the theory, i.e.,
\begin{align}
\mathcal{Z}(m_{q}) \equiv \int [d\bar{q}dq] \exp\left\{-S_{0}[\bar{q},q]  -m_{q}\int d^{4}x \, \bar{q}q(x)\right\}~.
\end{align}
The expressions for the two terms on the right hand side of Eq.~(\ref{eq:der_contribs}), read:
\begin{align}
\label{eq:der_val}
\frac{\partial \mathcal{O}(m_{q}) \rangle }{\partial m_{q}}\bigg|_{\rm val} &\equiv \frac{1}{\mathcal{Z}(m_{q})}\int [d\bar{q}dq]  \exp\left\{-S_{0}[\bar{q},q]  -m_{q}\int d^{4}x \, \bar{q}q(x)\right\} \frac{\partial}{\partial m_{q}} \mathcal{O}(\bar{q},q, m_{q}) \equiv \langle \frac{\partial\mathcal{O}(m_{q})}{\partial m_{q}}\rangle~,\\
\nonumber\\
\frac{\partial \mathcal{O}(m_{q})  }{\partial m_{q}}\bigg|_{\rm sea} &\equiv \frac{1}{\mathcal{Z}(m_{q})}\int [d\bar{q}dq] \frac{\partial}{\partial m_{q}} \left (\exp\left\{-S_{0}[\bar{q},q]  -m_{q}\int d^{4}x \, \bar{q}q(x)\right\} \right)  \mathcal{O}(\bar{q},q, m_{q}) - \frac{\partial\ln{\mathcal{Z}(m_{q})}}{\partial m_{q}} \langle \mathcal{O}(m_{q})\rangle \nonumber \\[8pt]
&= -\int d^{4}x \left[~\langle \bar{q}q(x)\,\mathcal{O}(m_{q})\rangle -\langle \bar{q}q(x)\rangle\langle \mathcal{O}(m_{q})\rangle~\right]\,.
\label{eq:der_sea}
\end{align}
Notice that the two contributions~(\ref{eq:der_val},\ref{eq:der_sea}) only involve the expectation values in the theory with a fixed sea-quark mass $m_{q}$, and therefore the derivative $\partial C_{q}(t)/ \partial m_{q}$ can be computed without having to generate new gauge configurations at different values of the sea-quark masses. In summary, the valence contribution to the derivative can be evaluated from the slope of the observable $\mathcal{O}(m_{q})$ w.r.t. the valence quark mass, while keeping the sea quark mass fixed. The sea quark contribution, instead, is given (up to a minus sign) by the gauge connected part of the correlation between the observable $\mathcal{O}(m_{q})$ and the spacetime integral of the scalar density $\bar{q}q(x)$.   

Bearing in mind Eqs.~(\ref{eq:der_val},\ref{eq:der_sea}) we can now discuss our calculation of $\partial C_{q}(t)/\partial m_{q}$. Given the smallness of the disconnected part, $C_{q}^{\rm disc}(t)$, of the full correlator $C_{q}(t)$, we neglect in our calculations its contribution to the derivative $\partial C_{q}/\partial m_{q}$, considering only the contribution from the dominant connected part.  We start by discussing the valence contribution to the derivative. If we denote by $m_{q}^{(L)}$ and $m_{q}^{(H)}$ the respective lowest and highest simulated valence quark masses for each flavor $q=u,d,s$, and by $C_{q}(t,m_{q})$ the Euclidean correlator evaluated with a given valence $m_{q}$ (at the fixed simulated values of the sea-quark masses), we determine the valence contribution as,
\begin{align}
\label{eq:der_val_num}
\frac{\partial C_{q}(t)}{\partial m_{q}}\bigg|_{\rm val} \simeq \frac{ C_{q}(t,m_{q}^{H}) - C_{q}(t,m_{q}^{L})}{ \Delta m_{q} } + \mathcal{O}(\Delta m_{q})~,\qquad \Delta m_{q} \equiv m_{q}^{H} - m_{q}^{L}~.
\end{align}
The simulated valence quark masses are ensured to satisfy $|\Delta m_{q}| \lesssim 0.1\, m_{q}$ of the average mass, in both situations $q=u,d$ and $q=s$, cf. Tables~\ref{tab:simudetails}-\ref{tab:masses}. Clearly, our evaluation of the valence contribution to the derivative using Eq.~(\ref{eq:der_val_num}) is acceptable only if $\Delta m_{q}$ is sufficiently small so that the higher order corrections to Eq.~(\ref{eq:der_val_num}) are indeed negligible. We tested the validity of this assumption on a single ensemble (B64), and in the (more relevant) case of the strange quark contribution, by considering in addition to the valence masses given in Table~\ref{tab:masses}) two more values, namely, $am_{s}=0.018, 0.020$. We evaluated the derivative using all four masses and found that the typical error committed by estimating the derivative using Eq.~(\ref{eq:der_val_num}) is about $2\%$, thus similar in size as our (total) statistical error of $\overline{H}_{s}(0)$. For that reason we attribute an additional $2\%$ of systematic uncertainty to the strange quark mass derivative of the correlator evaluated using Eq.~(\ref{eq:der_val_num}). In the case of the light quark contribution, instead, no additional systematic uncertainty is added, given that the contribution of the derivative for $q=u,d$ is approximately one order of magnitude smaller than for $q=s$, and this systematic effect is much smaller than the statistical error of our results.

In the case of the sea quark contribution to the derivative, we make use of Eq.~(\ref{eq:der_sea}) and of our determination of the light and strange scalar densities, to compute $\partial C_{q}(t)/ \partial m_{q}\big|_{\rm sea}$ using, 
\begin{align}
\label{eq:der_sea_num}
\frac{\partial C_{q}(t) \rangle }{\partial m_{q}}\bigg|_{\rm sea} = - \int d^{4}x \,\biggl[~\langle \bar{q}q(x)\,C_{q}(t)\rangle -\langle \bar{q}q(x)\rangle\langle C_{q}(t)\rangle~\biggr]~, \qquad q=u, d, s~.
\end{align}
It is important to notice that the product $m_{q}\partial/\partial m_{q}$ entering Eq.~(\ref{eq:subtraction_procedure}) is Renormalization Group Invariant (RGI), hence Eqs.~(\ref{eq:der_val_num})-(\ref{eq:der_sea_num}) can be evaluated using the bare quark masses and the bare scalar densities.

 As it will be detailed in the numerical Section, the subtraction of the log-derivative in Eq.~(\ref{eq:subtraction_procedure}) has a much higher impact in the case of the strange-quark, where the contribution is enhanced due to the large quark-mass ratio $m_{s}/m_{\ell} \sim 27$ w.r.t. the case of the light quarks. In practice, it turns out that the subtraction is an order $5\%$ effect in the case $q=u,d$, while it is of about $30-40\%$ in the case $q=s$.
\begin{center}
\begin{table}[]
\renewcommand{\arraystretch}{2.}
    \centering
    \begin{tabular}{| c || c |  c ||  c |}
    \hline
        ~~ Ensemble ~~ & ~~ $am_s^L$ ~~ & ~~ $am_s^H$ ~~ & $  am_{\ell}' $ ~~ \\
        \hline
        B64 & $0.019$ & $0.021$ & $0.0006675$ \\ \hline
        B96 & $0.019$ & $0.021$ & $0.0006675$ \\
        \hline
        C80 & $0.016$ & $0.018$ & $0.000585$ \\
        \hline
        D96 & $0.014$ & $0.015$ & $0.0004964$  \\
        \hline
    \end{tabular}
    \caption{\it \small Values of the two bare strange-quark masses, $a m_{s}^{L}$ and $a m_{s}^{H}$, and of the valence light quark mass $am_{\ell}'$ (see text for details), for each of the four ensembles of Table\,\ref{tab:simudetails}. }
    \label{tab:masses}
\end{table}
\end{center}

\section{Numerical results}
\begin{figure}
    \centering
    \includegraphics[scale=0.37]{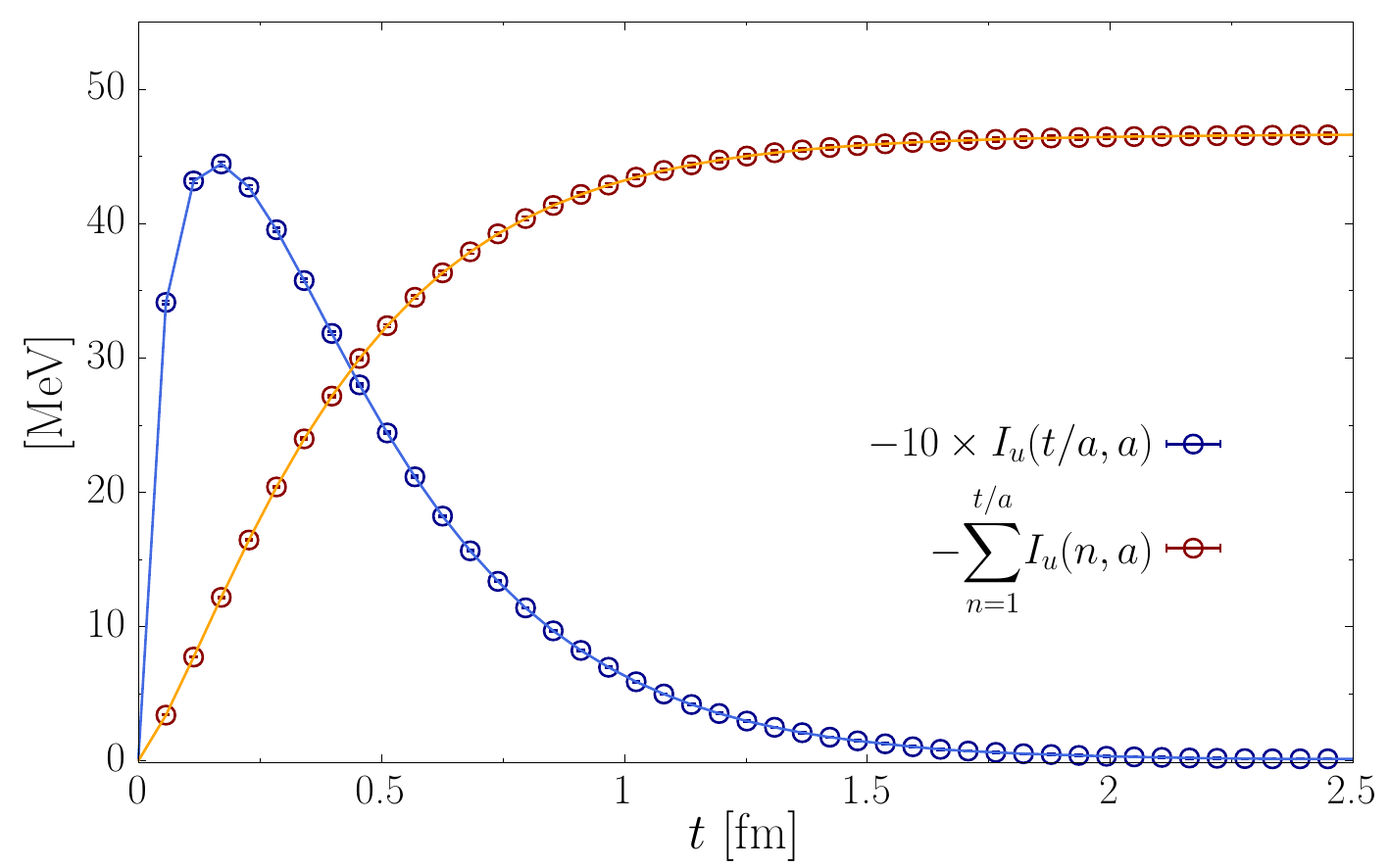} 
    \includegraphics[scale=0.37]{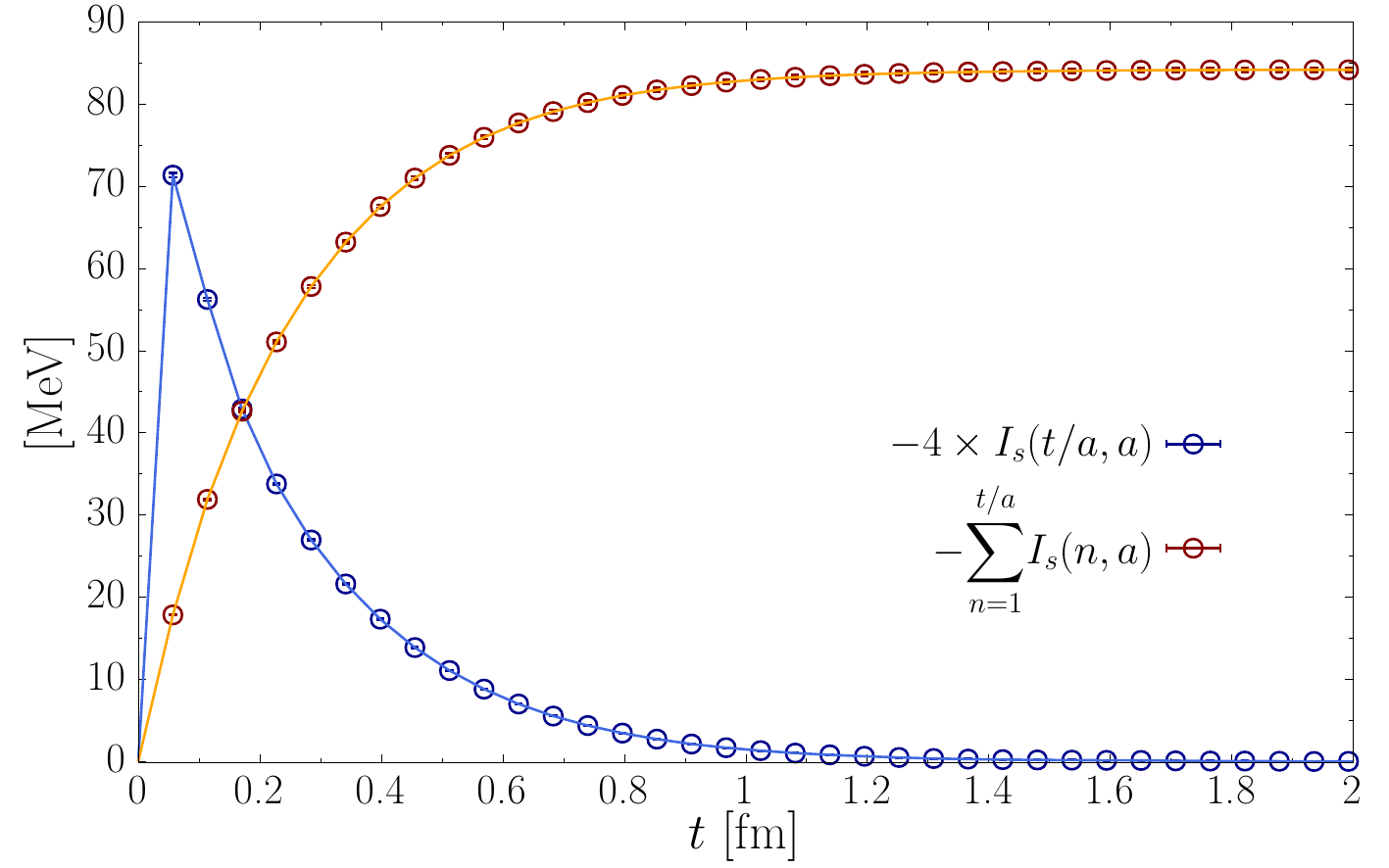}
    \caption{\small\it The values of $I_{q}(n,a)$ (blue data points) and of its partial sums (red data points) on our finest lattice spacing ensemble and for the ``tm'' regularization. The left and right panels correspond to the cases $q=u$ and $q=s$, respectively. For the strange-quark contribution the results have been already interpolated to the physical strange quark mass. }
    \label{fig:1}
\end{figure}

\begin{figure}
    \centering
    \includegraphics[scale=0.37]{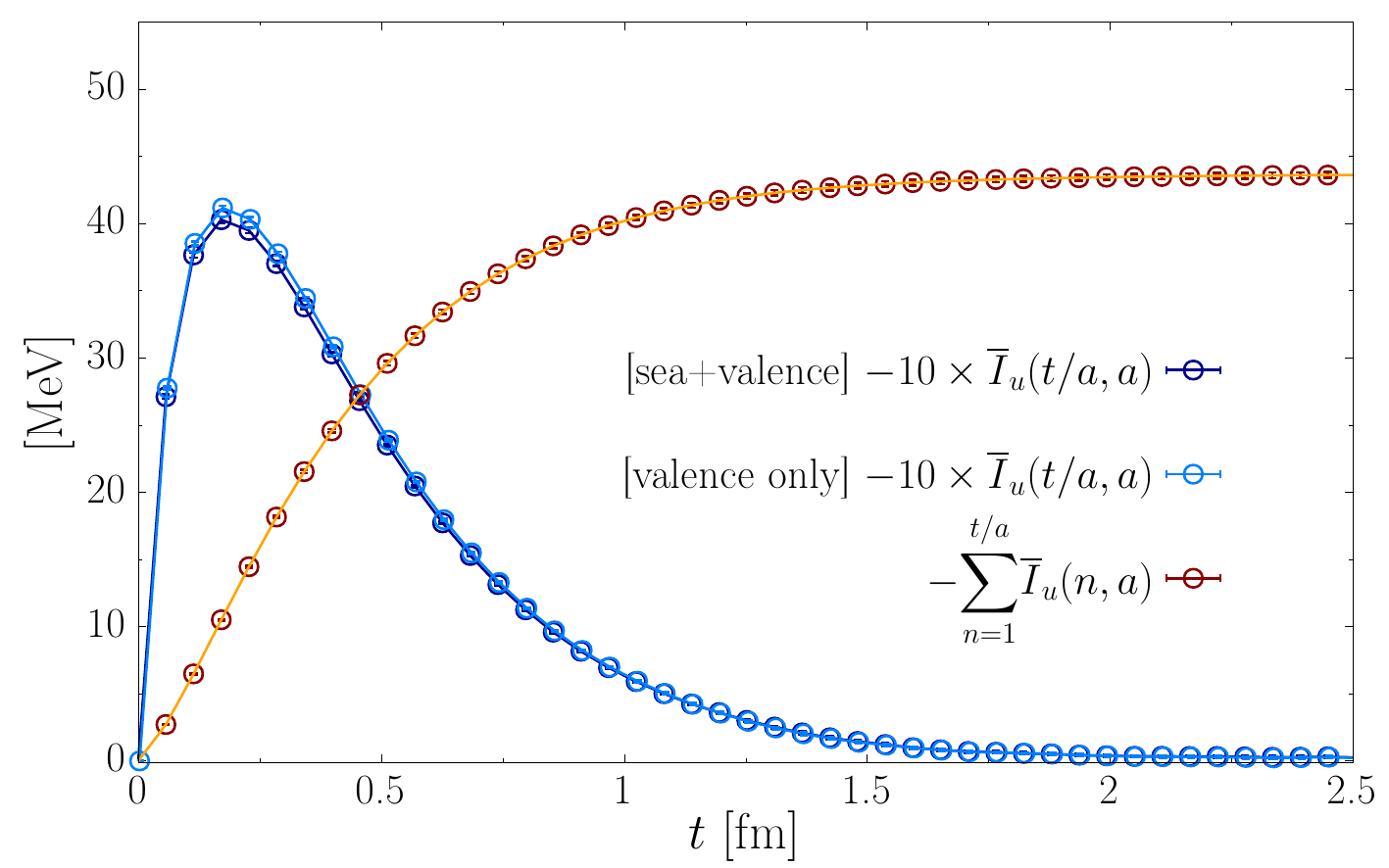} 
    \includegraphics[scale=0.37]{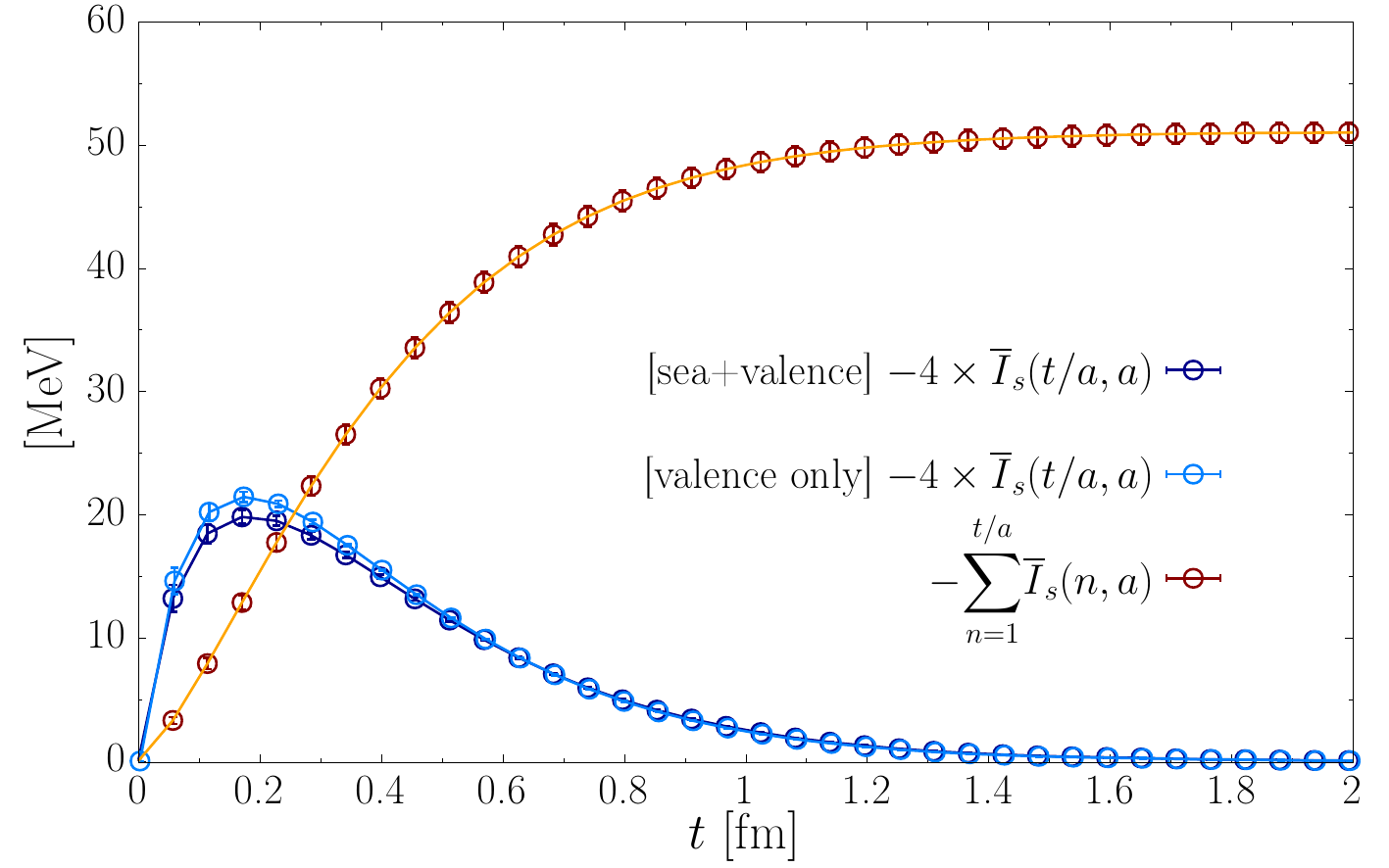}
    \caption{\small\it The values of $\overline{I}_{q}(n,a)$ (blue data points) and of its partial sum (red data points) on our finest lattice spacing ensemble and for the ``tm'' regularization. The left and right figures correspond to the cases $q=u$ and $q=s$, respectively. For the strange-quark contribution the results have been already interpolated to the physical strange quark mass. The data points in light-blue correspond to the value of $\overline{I}_{q}(n,a)$ obtained neglecting the sea-quark contribution in Eq.~(\ref{eq:der_sea_num}) to the quark-mass derivative of the correlator.}
    \label{fig:2}
\end{figure}

We now turn to the results of our lattice calculation of $\overline{H}_{q}(0)$ for $q=u,d,s$. At finite lattice spacing we evaluate the (unsubtracted) form factor $H_{q}(0)$ from the knowledge of the correlator $C_{q}(na)$, with $n\in \{1, \ldots, T/(2a)\}$, where $T$ is the physical temporal lattice extent. We use the following discretized version of the integral~(\ref{eq:kernel}),
\begin{align}
H_{q}(0) =  \sum_{n=1}^{n_{max}} I_{q}(n,a) \equiv \sum_{n=1}^{n_{max}}\, 2a^{2}n\cdot C_{q}(na)~, 
\end{align}
where $n_\mathrm{max} \leq T/2a$ is chosen large enough so that the sum in the previous equation already converged to its $n_\mathrm{max}\to\infty$ limit within errors. We optimize the value of $n_\mathrm{max}$ so as to exclude from  integration the contribution from the very large Euclidean times which only increase the noise in the resulting $H_{q}(0)$, but give no contribution to its signal.  
In the plots shown in Figure~\ref{fig:1} we illustrate the behavior of $I_{q}(n,a)$ and of its partial sums, for a selected gauge ensemble, the ``tm" regularization, and for both the up- and the strange-quark contributions~\footnote{The down-quark contribution is very similar to the up-quark one. Their difference is proportional to the disconnected diagram in Figure~\ref{fig:Feyn}, which is two orders of magnitude smaller than the connected part.} 
As it can be appreciated from the plots, the integrand $I_{q}(t/a,a)$ is peaked at very short time separations $t$, of the order of the lattice spacing, with the peak being more pronounced in the case of the strange quark. This is a manifestation of the logarithmic divergence affecting $H_{q}(0)$, which is expected to be more sizable for heavier quarks, and which we subtract away by using the prescription described above and specified in Eq.~(\ref{eq:subtraction_procedure}).
To this purpose, we evaluate on each gauge ensemble the mass-derivative of the correlator $C_{q}(t)$ for $q=u,d,s$, using the procedure discussed in Section~\ref{sec:derivative}.
We then compute the subtracted form-factor $\bar{H}_{q}(0)$ as,
\begin{align}
\overline{H}_{q}(0) = {\displaystyle \sum_{n=1}^{n_{max}}} \overline{I}_{q}(n,a) \equiv {\displaystyle \sum_{n=1}^{n_{max}}} 2a^{2}n\cdot\left[ C_{q}(na) - m_{q}\frac{\partial C_{q}(na)}{\partial m_{q}}\right]~.
\end{align}
The subtracted counterpart of Figure~\ref{fig:1} is shown in Figure~\ref{fig:2}. 
The effect of the subtraction amounts to a $\mathcal{O}(5\%)$ shift of the form factor in the case of the up- and down-quark, while for the strange-quark, as expected, such an effect is much larger and the subtracted form factor $\overline{H}_{s}(0)$ is $\simeq 30-40\%$ smaller than the unsubtracted one, $H_{s}(0)$. In Figure~\ref{fig:2} we show separately the value of $\overline{I}_{q}(n,a)$ obtained by neglecting the sea quark contribution to the derivative (Eq.~(\ref{eq:der_sea_num})). 
While the sea quark contribution to the derivative is smaller than the valence one (by about one order of magnitude in the case of the strange-quark) it is non-negligible w.r.t. our current statistical accuracy. For that reason both terms in Eq.~\eqref{eq:der_contribs} should be and are taken into account. The results for the unsubtracted and subtracted form factors, ${H}_{q}(0)$ and $\overline{H}_{q}(0)$, for both regularization schemes and for $q=u,d,s$ are presented in Table~\ref{tab:ffs}.

\begin{center}
\renewcommand{\arraystretch}{2.}
\begin{table}[]
    \centering
    \begin{tabular}{|c| c| c  c  c  | c   c   c| }
      \hline
     Ensemble & Form factor [MeV] & tm ($q=u$) & tm ($q=d$)  & tm ($q=s$) & OS ($q=u$) & OS ($q=d$) & OS ($q=s$) \\
     \hline
     B64 & $-{H}_{q}(0)$ &~ 46.59(15) ~  & ~ 47.29(20) ~ &  ~ 78.98(18) ~   & ~ 45.86(14) ~ & ~ 46.58(20)  ~ & ~ 78.61(18) ~   \\
          & $-\overline{H}_{q}(0)$ &~ 43.65(21) ~  & ~ 44.35(28) ~ &  ~ 51.01(75) ~   & ~ 42.84(19)  ~ & ~ 43.56(26)    ~ & ~ 50.69(74)   ~   \\ \hline 
     B96 & $-{H}_{q}(0)$& ~ 46.79(14) ~ & ~ 47.51(19) ~  & ~ 78.97(17) ~ & ~  46.05(13) ~ & ~ 46.76(19)  ~ & ~ 78.59(17) ~  \\
         & $-\overline{H}_{q}(0)$& ~ 43.88(20) ~ & ~ 44.60(28) ~  & ~ 50.97(86) ~ & ~  43.28(23) ~ & ~ 43.99(30)   ~ & ~ 50.64(85)   ~  \\ \hline 
     C80 & $-{H}_{q}(0)$& ~ 46.23(17) ~ & ~ 47.03(19) ~ & ~ 81.00(18) ~   & ~ 45.72(16) ~ & ~ 46.52(19)   ~ & ~  80.72(17)  ~  \\
         & $-\overline{H}_{q}(0)$& ~ 43.44(33) ~ & ~ 44.24(30) ~ & ~ 50.18(86) ~   & ~ 43.31(28) ~  & ~ 44.12(28)   ~ & ~ 50.02(84)   ~ \\ \hline 
     D96 & $-{H}_{q}(0)$ &~ 46.68(19) ~ & ~ 47.38(24) ~ & ~  84.21(25) ~ & ~ 46.28(20) ~ & ~ 46.98(23)   ~ & ~ 84.00(24) ~ \\
          & $-\overline{H}_{q}(0)$ &~ 43.69(24) ~ & ~ 44.38(28) ~ & ~  51.10(80) ~ & ~ 43.35(26) ~ & ~ 44.05(28)   ~ & ~ 50.95(79)   ~ \\
     \hline
    \end{tabular}
    \caption{\small\it Results of the unsubtracted [$H_{q}(0)$] and subtracted [$\overline{H}_{q}(0)$] form factors obtained for each of the ensembles of gauge field configurations specified in Table~\ref{tab:simudetails}, for both the light non-strange $q = u, d$ and the strange quark case $q=s$. As described in the text, we use two kinds of regularization schemes, denoted by ``tm" and ``OS", and the form factor results are provided for both regularizations. }
    \label{tab:ffs}
\end{table}
\end{center}

We now move to the discussion of extrapolation of our lattice results both to the continuum ($a\to 0$) and to the infinite volume ($L\to \infty$).

\subsection{Continuum and infinite-volume extrapolation}
\label{sec:cont_vol}
\begin{figure}
    \centering
    \includegraphics[scale=0.37]{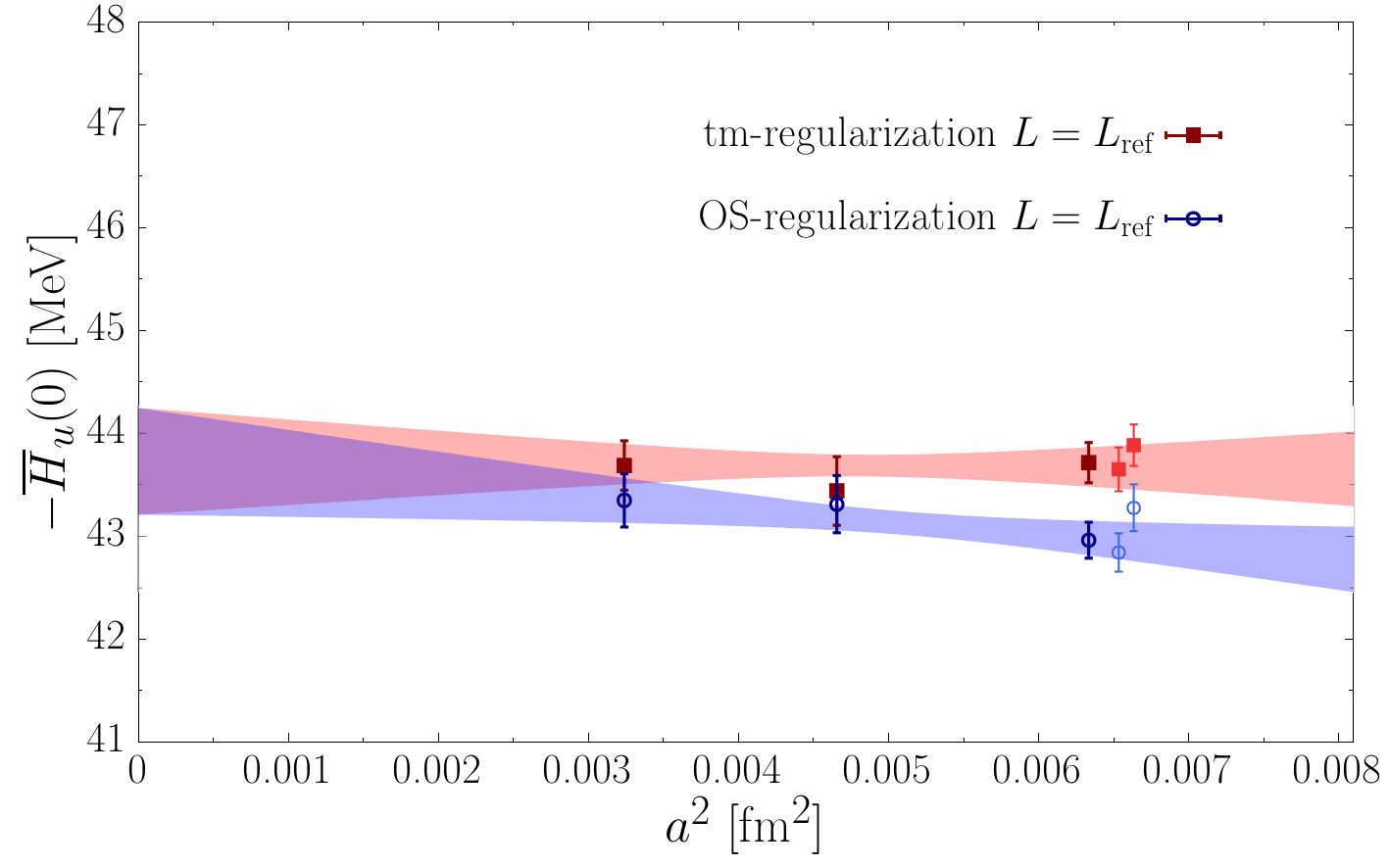}   \includegraphics[scale=0.37] {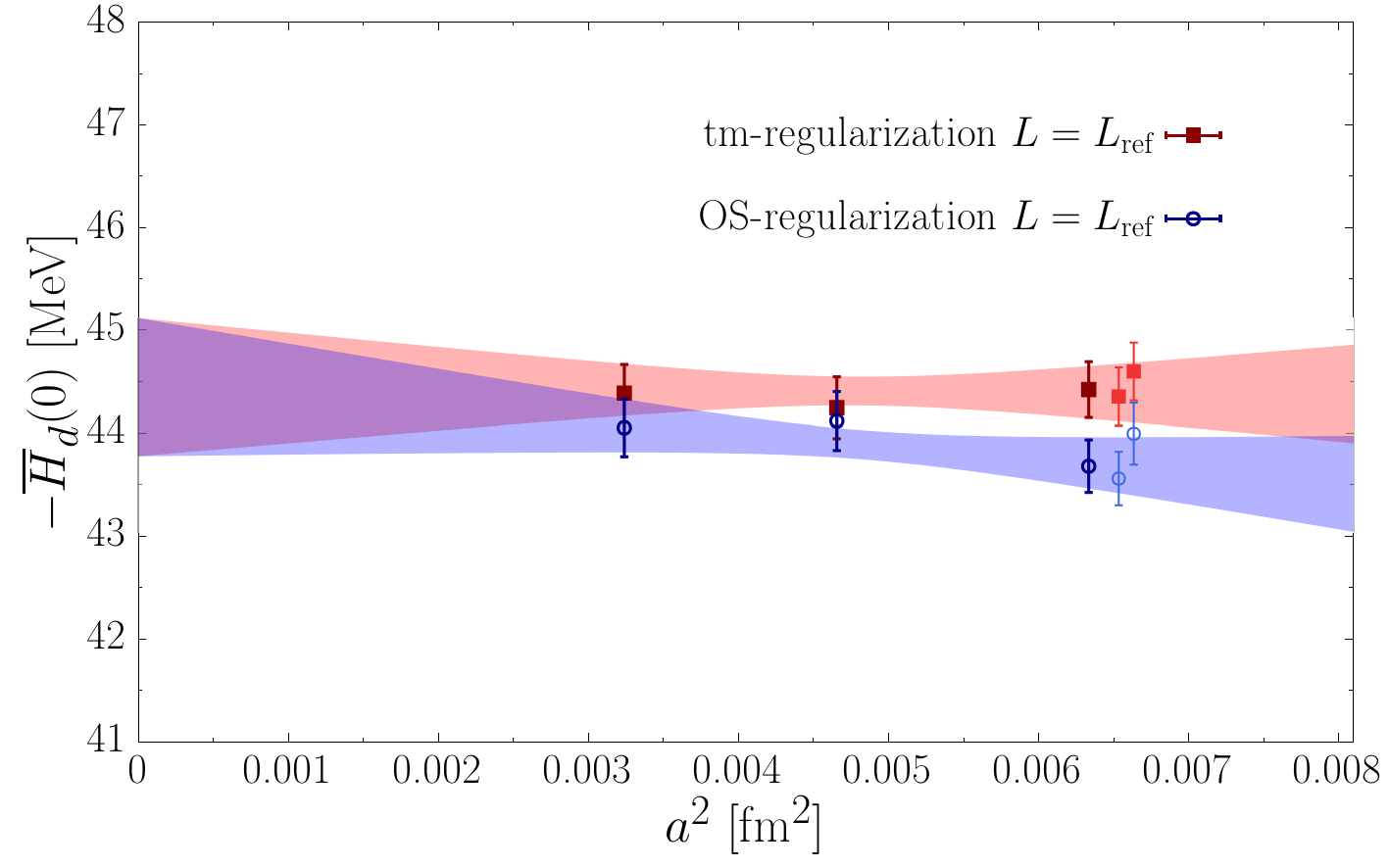}
    \\ \includegraphics[scale=0.37]{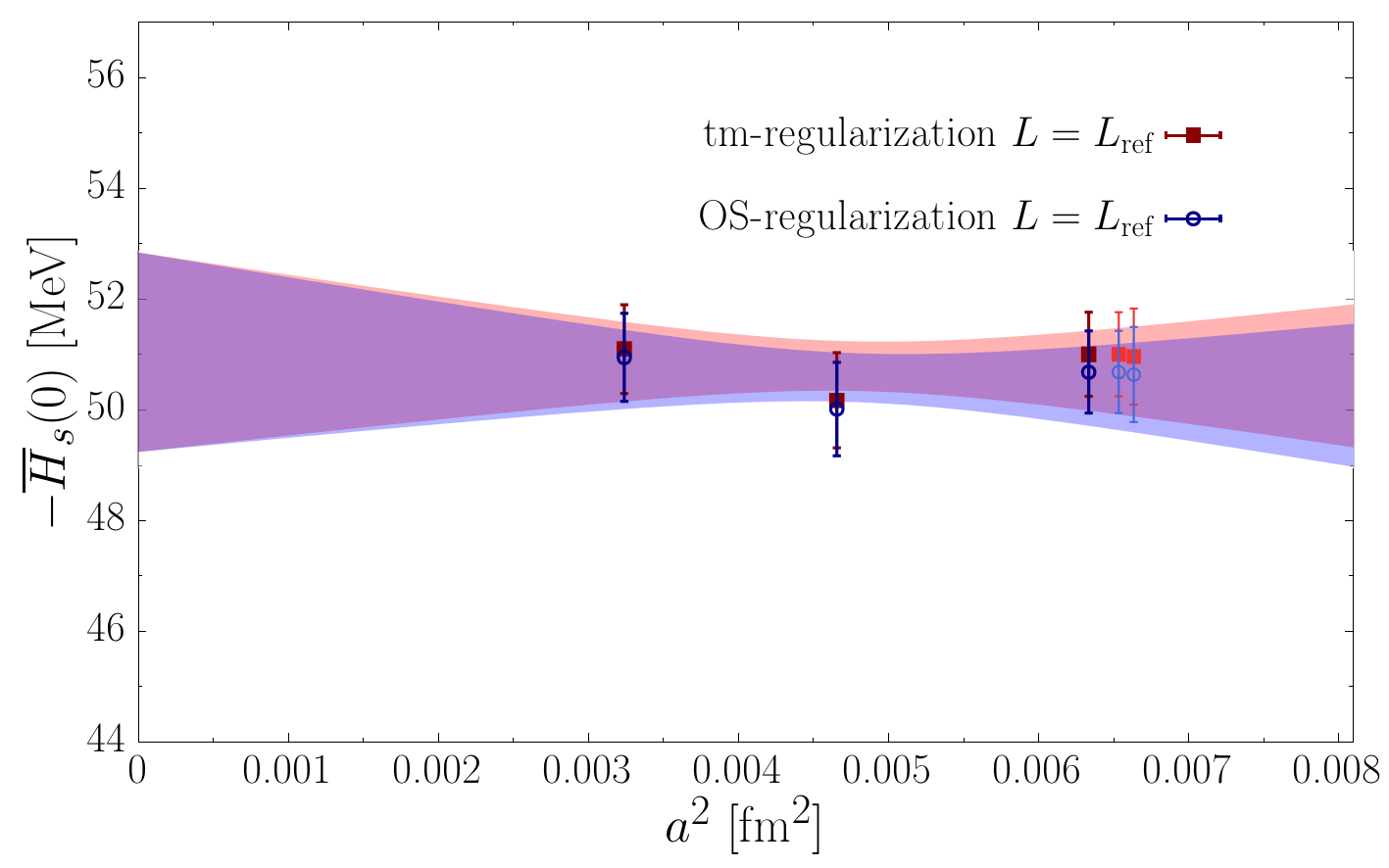}
    \caption{\small\it Continuum limit extrapolation of $\overline{H}_{u}(0)$ (top-left panel), $\overline{H}_{d}(0)$ (top-right panel),  and $\overline{H}_{s}(0)$ (bottom panel). The dark red and dark blue data points correspond to our results at $L=L_{\rm ref}$ obtained by respectively using the `tm'' and ``OS'' regularization. In order to highlight the typical size of the FSE's we also show the raw data of the ensembles B64 and B96 (light red and light blue data points) at $a\simeq 0.079~{\rm fm}$. Finally, the coloured bands correspond to the best fit functions obtained in the continuum limit extrapolation using the ansatz~(\ref{eq:cont_ansatz}).  }
    \label{fig:3}
\end{figure}

As described in the previous Sections, we evaluated the (subtracted) form factors $\overline{H}_{u,d,s}(0)$ on the ensembles listed in  Table~\ref{tab:simudetails} by using the 
``tm'' and the ``OS'' regularizations. Apart from B96, all the ensembles from Table~\ref{tab:simudetails} have very similar spatial volumes. The ensembles C80 and D96 have (within uncertainties) the lattice extent $L\approx 5.46~{\rm fm}$, while for the ensemble B64 it is slightly shorter, $5.09~{\rm fm}$. For the ensemble B96, instead, it is much larger, i.e. more than $L \approx 7.5~{\rm fm}$. Our strategy to perform the combined extrapolation to the continuum and infinite volume limit can be summarized as follows. From the knowledge of the form factors on the ensembles  B64 and B96 we interpolate $\overline{H}_{q}(0)$ at $\beta=1.778$ to the reference lattice extent of $L_{\rm ref}=5.46~{\rm fm}$. The interpolation is performed separately for the two regularizations  (``tm'' and ``OS''), using a linear function in $\exp{-M_{\pi}L}$. In this way we can perform the continuum extrapolation at fixed volume $V_{\rm ref}= L_{\rm ref}^{3}$.  We then associate to our  extrapolated results at $V_{\rm ref}$ a systematic uncertainty $\Delta_{L}\overline{H}_{q}$ due to finite size effects (FSE's), which we estimate as
\begin{align}
\label{eq:FSEs}
\Delta_{L} \overline{H}_{q} = \max_{r={\rm{tm}, \rm{OS}}} \left\{ \Delta_{q}^{r} \erf\left(\frac{\Delta_{q}^{r}}{\sqrt{2}\sigma_{\Delta_{q}^{r}}}\right)    \right\}~, \qquad q=u,d,s~,
\end{align}
where we have defined
\begin{align}
\Delta_{q}^{r} = \big| \overline{H}_{q}^{r}(0, B96) - \overline{H}_{q}^{r}(0, B64) \big| ~,
\end{align}
and $\sigma_{\Delta_{q}^{r}}$ is its statistical uncertainty.
%\begin{align}
%\Delta_{L} \overline{H}_{q} = \max_{r=%{\rm{tm}, \rm{OS}}}\left\{ \big| \overline{H}_{q}^{r}(0, B96) - \overline{H}_{q}^{r}(0, B64) \big|   \right\}~,\qquad q=\ell,s~.
%\end{align}
The continuum extrapolation at $V=V_{\rm ref}$ is carried out through a combined  fit of the two regularizations  (``tm'' and ``OS'') to the common continuum limit result, using the linear ansatz in $a^{2}$
\begin{align}
\label{eq:cont_ansatz}
\overline{H}^{\rm tm}_{q}(0, a^{2}) = A_{q} + B_{q}^{\rm tm} a^{2}~, \qquad \overline{H}^{\rm OS}_{q}(0, a^{2}) = A_{q} + B_{q}^{\rm OS} a^{2}~,\qquad q=u,d,s~,
\end{align}
where $A_{q}, B_{q}^{\rm tm}$ and $B_{q}^{\rm OS}$ are the free fit parameters. We have minimized $\chi^{2}$ which properly takes into account the correlation between the data obtained with ``tm'' and ``OS'' regularizations at the same value of $\beta$. The results of the combined continuum and infinite volume extrapolations are shown in the plots shown in Figure~\ref{fig:3}. The red and blue data points correspond to the results obtained with two regularizations at $L=L_{\rm ref}$, while the red and blue bands correspond to the best fit functions with a common continuum limit value. For both $\overline{H}_{u}(0)$, $\overline{H}_{d}(0)$ and $\overline{H}_{s}(0)$ the $\chi^{2}/N_{\rm dof}$ of the fit is $0.6-0.7$ with $N_{\rm dof}=3$. In order to highlight the typical size of the FSE's we show in the same figure the raw data obtained on the B64 and B96 ensembles at $a \simeq 0.079$~fm. They are shown as light red and light blue data points for the  ``tm'' and ``OS'' regularizations,  respectively. A few comments are in order: First of all, the observed FSE's are very small, completely negligible (within errors) for $\overline{H}_{s}(0)$, however similar in size to our statistical error for $\overline{H}_{u}(0)$ and $\overline{H}_{d}(0)$. Furthermore, the typical size of the cutoff effects is very modest and at most of the order of a few percent for both contributions. As it can be seen from the plots, the two regularizations give rise to very similar results already at our simulated values of the lattice spacing. This is different from what has been observed in the case of the correlation function of two electromagnetic currents,  $\langle J_{\rm em} J_{\rm em}^{\dag}\rangle$, which enters the lattice calculation of the leading order hadronic vacuum polarization contribution to $(g-2)_\mu$~\cite{Alexandrou:2022amy}. Our qualitative explanation of this feature is that, as shown by our tree level calculation in  Appendix~\ref{appendix_A}, the ``tm'' and ``OS'' tensor-vector correlation functions coincide to all orders in $\mathcal{O}(\alpha_{s}^{0} a^{n} m_{q}^{r})$, and they only differ at order $\mathcal{O}(\alpha_{s})$. This is not true for the vector-vector correlation function (see Appendix~E of Ref.~\cite{Alexandrou:2022amy}). 
In the continuum and the infinite-volume limit, our final results in the $\overline{\mathrm{MS}}$ scheme, read:
\begin{align}
\overline{H}_{u}(0,  2~{\rm GeV}) &= f_{\gamma,u}^{\perp}(2~{\rm GeV})   = -43.73(52)_{L_{\rm ref}}(38)_{\rm FSE} ~{\rm MeV} = -43.73(64)~{\rm MeV}~, \\[8pt]
\overline{H}_{d}(0,  2~{\rm GeV}) &= f_{\gamma,d}^{\perp}(2~{\rm GeV})   = -44.45(67)_{L_{\rm ref}}(32)_{\rm FSE} ~{\rm MeV} = -44.45(74)~{\rm MeV}~, \\[8pt]
\overline{H}_{s}(0,  2~{\rm GeV}) &= f_{\gamma,s}^{\perp}(2~{\rm GeV}) = -51.0(1.8)_{L_{\rm ref}}~(0)_{\rm FSE} ~{\rm MeV} \,\, = -51.0(1.8)~{\rm MeV}~,
\end{align}
where the first error corresponds to the total (statistical and systematic) error obtained after performing the continuum extrapolation at $L=L_{\rm ref}$, and the second one to the estimate of the FSE's obtained by using Eq.~(\ref{eq:FSEs}). 
Our results can be compared with those of Ref.~\cite{Bali:2020bcn}, namely $f_{\gamma,u}^{\perp}(2~{\rm GeV}) \simeq f_{\gamma,d}^{\perp}(2~{\rm GeV})   = -45.4(1.5)~{\rm MeV}$ and $f_{\gamma,s}^{\perp}(2~{\rm GeV}) = -68(5)~{\rm MeV}$. While we find a good agreement with Ref.~\cite{Bali:2020bcn} for $f_{\gamma,u/d}^{\perp}$, we disagree by several standard deviations for $f_{\gamma,s}^{\perp}$. It must be noted that, in contrast to what is done in this work, the strange quark mass derivative of the form-factor $H_s(0)$, which is needed in order to cancel the logarithmic divergence of the unsubtracted form factor~\eqref{eq:subtraction_procedure}, has been estimated indirectly in Ref.~\cite{Bali:2020bcn}, from the dependence of the form factor $H_{u/d
} (0)$ on the light quark mass $m_{\ell}$. Besides the observed discrepancy in the strange quark case, we obtain that the resulting uncertainties in both $\overline {H}_{u/d}(0)$ and $\overline {H}_{s}(0)$ are smaller than those quoted in Ref.~\cite{Bali:2020bcn} by a factor $2.3$ and $2.8$, respectively. 
That gain in precision does not translate to the precision of the value of susceptibility of the quark condensate. The reason is that the current value of $\langle \bar q q \rangle $ as extracted from the simulations with $N_f=2+1$ dynamical flavors,
$\langle \bar q q\rangle= - (272(5)~\mev)^3$~\cite{FlavourLatticeAveragingGroupFLAG:2021npn}, is more accurate than the one obtained from simulations with $N_f=2+1+1$, $\langle \bar q q\rangle= - (286(23)~\mev)^3$~\cite{FlavourLatticeAveragingGroupFLAG:2021npn}. By using the latter value we obtain 
$\chi_u^{\msbar}(2\,\mathrm{GeV})= 1.87(45)~\gev^{-2}$ and $\chi_d^{\msbar}(2\,\mathrm{GeV})= 1.90(46)~\gev^{-2}$. However, since the inclusion of the charm quark in the sea is very unlikely to affect the value of the quark condensate obtained from simulations with $N_f=2+1$, we may as well use the $\mathrm{MS}$ value obtained from simulations with $N_f=2+1$ at $\mu=2$~GeV, $\langle \bar q q\rangle = - (272(5)~\mathrm{MeV})^3$~\cite{FlavourLatticeAveragingGroupFLAG:2021npn}, which then leads to  $\chi_u^{\msbar}(2\,\mathrm{GeV}) = 2.17(12)~\gev^{-2}$ and $\chi_d^{\msbar}(2\,\mathrm{GeV})=2.21(13)~\gev^{-2}$.

\section{Conclusions}

In this work we computed the normalization of the photon's leading twist DA, $f_{\gamma,q}^{\perp}(0)$, by means of numerical simulations of QCD with $N_f=2+1+1$ dynamical quarks on the lattice. We rely on the gauge field configurations generated by the ETM Collaboration on which we compute the desired two-point correlation functions. We included the disconnected diagrams in our computation even though we verified that their contribution is orders of magnitude smaller than the connected ones. The additive and multiplicative renormalization is carried out fully nonperturbatively and then converted to the $\msbar$ scheme by using the $4$-loop perturbative matching.

By using two various lattice regularizations (denoted as ``tm" and ``OS" in the text) we were able to better control the extrapolation to the continuum limit. After a careful assessment of systematic uncertainties due to continuum extrapolation ($a\to 0$) and to the physical volume ($L\to \infty$) we find that the normalization of the leading twist DA of a photon coupling to $u$ or $d$ quark, in the $\overline{\mathrm{MS}}$ renormalization scheme, is 
\begin{align}
f_{\gamma,u}^{\perp}(2~{\rm GeV})   &= -43.73(64)~{\rm MeV}~, \\[8pt]
f_{\gamma,d}^{\perp}(2~{\rm GeV})   &= -44.45(74)~{\rm MeV}~,
\end{align}
while the coupling to the strange quark is,
\begin{align}
&f_{\gamma,s}^{\perp}(2~{\rm GeV})   = -51.0(1.8)~{\rm MeV}~.
\end{align}
Both results represent improvement with respect to the previous lattice QCD estimates. 
It was shown in Ref.~\cite{Bali:2020bcn} that $f_{\gamma,u/d}^{\perp}$ are indistinguishable from their values in the chiral limit, $\displaystyle{\lim_{m_\ell\to 0}} f_{\gamma,u/d}^{\perp}$. One can therefore deduce the value of  the 
magnetic susceptibility of the quark condensate, $\chi_{d} \simeq \chi_{u} = f_{\gamma,u}^{\perp}/\langle \bar{q} q\rangle = 2.17(12)~\gev^{-2}$, where we used the FLAG value of $\langle \bar{q} q\rangle^{\msbar}(2~\gev)$ obtained from simulations with $N_f=2+1$~\cite{FlavourLatticeAveragingGroupFLAG:2021npn}. One would eventually prefer to use the condensate value from simulations with $N_f=2+1+1$ but its current value is much less accurate, although its value is unlikely to change with respect to the $N_f=2+1$ result.

\section{Acknowledgements}
We thank all the members of the ETM Collaboration for the most enjoyable collaboration, the authors of Ref.~\cite{Ball:2002ps} for correspondence, and of Ref.~\cite{Bali:2020bcn} for comments on the manuscript. We gratefully acknowledge CINECA and EuroHPC JU for giving this project access to Leonardo supercomputing hosted at CINECA.  We acknowledge CINECA for the provision of GPU time under the specific
initiative INFN-LQCD123 and IscrB\_S-EPIC. F.S. and G.G are supported by the Italian Ministry
of University and Research (MIUR) under grant PRIN20172LNEEZ. F.S. and G.G. are supported by
INFN under GRANT73/CALAT. F.S. is supported by ICSC – Centro Nazionale di Ricerca in High Performance
Computing, Big Data and Quantum Computing, funded by European Union –
NextGenerationEU. S.B. has received funding from the European High-Performance Computing Joint Undertaking (JU) under grant 
agreement No 101118139. The JU receives support from the European Union’s Horizon Europe Programme. 
The authors gratefully acknowledge the Gauss Centre for Supercomputing e.V. (www.gauss-centre.eu) for funding this project by providing computing time on the GCS Supercomputers SuperMUC-NG at Leibniz Supercomputing Centre and JUWELS~\cite{JUWELS} at Juelich Supercomputing Centre.
The authors acknowledge the Texas Advanced Computing Center (TACC) at The University of Texas at Austin for providing HPC resources (Project ID PHY21001).
The authors gratefully acknowledge PRACE for awarding access to HAWK at HLRS within the project with Id Acid 4886.
We acknowledge the Swiss National Supercomputing Centre (CSCS) and the EuroHPC Joint Undertaking for awarding this project access to the LUMI supercomputer, owned by the EuroHPC Joint Undertaking, hosted by CSC (Finland) and the LUMI consortium through the Chronos programme under project IDs CH17-CSCS-CYP and CH21-CSCS-UNIBE as well as the EuroHPC Regular Access Mode under project ID EHPC-REG-2021R0095.
The open-source packages tmLQCD~\cite{Jansen:2009xp,Abdel-Rehim:2013wba,Deuzeman:2013xaa,Kostrzewa:2022hsv}, LEMON~\cite{Deuzeman:2011wz}, DD-$\alpha$AMG~\cite{Frommer:2013fsa,Alexandrou:2016izb,Bacchio:2017pcp,Alexandrou:2018wiv},QPhiX~\cite{joo2016optimizing,Schrock:2015gik} and QUDA~\cite{Clark:2009wm,Babich:2011np,Clark:2016rdz} have been used in the ensemble generation.

\bibliography{biblio.bib}
\bibliographystyle{JHEP}

\appendix

\section{Tree-level calculation of $C_{q}(t)$}
\label{appendix_A}
In this appendix we present the calculation of the free-theory correlator on the lattice, and obtain the relation appearing in Eq.~(\ref{eq:log_divergence}).  In the twisted-mass regularization, the momentum-space free-theory propagator of a quark-field $\psi_{q}$ of mass $m_{q}$ is given by
\begin{align}
\label{eq:prop_mom}
\langle \psi_{q}(p)\bar{\psi}_{q}(-p)\rangle = \frac{-i\gamma_{\mu}\tilde{p}_{\mu} +m_{q} +ir\gamma_{5}\frac{a}{2}\sum_{\mu}\hat{p}_{\mu}^{2}}{ \sum_{\mu} \tilde{p}_{\mu}^{2}  + 
m_{q}^{2} + \frac{a^{2}}{4}\left( \sum_{\mu}\hat{p}_{\mu}^{2}\right)^{2} }  ~,
\end{align}
where $r=\pm 1$ is the sign of the twisted Wilson term, the $\gamma^{\mu}$
are here the Euclidean gamma matrices ($\{\gamma^{\mu},\gamma^{\nu}\} = 2\delta^{\mu\nu}$), and 
\begin{align}
\tilde{p}_{\mu} = \frac{1}{a}\sin{(ap_{\mu})},\qquad \hat{p}_{\mu} = \frac{2}{a}\sin{(\frac{ap_{\mu}}{2})}~. \end{align}
The coordinate-space quark propagator is then given by ($x_{0}-y_{0} = t$)
\begin{align}
\label{eq:prop_coord}
\langle \psi_{q}(x)\bar{\psi}_{q}(y)\rangle = \int_{-\frac{\pi}{a}}^{\frac{\pi}{a}} \frac{dp_{0}}{2\pi} \int_{-\frac{\pi}{a}}^{\frac{\pi}{a}} \frac{d^{3}\bm{p}}{(2\pi)^{3}}e^{ip_{0}t} e^{i\bm{p}\cdot(\bm{x-y})}\cdot\frac{-i\gamma_{\mu}\tilde{p}_{\mu} +m_{q} +ir\gamma_{5}\frac{a}{2}\sum_{\mu}\hat{p}_{\mu}^{2}}{ \sum_{\mu} \tilde{p}_{\mu}^{2} + m_{q}^{2} + \frac{a^{2}}{4}\left( \sum_{\mu}\hat{p}_{\mu}^{2}\right)^{2} }~,
\end{align}
The integral over $p_{0}$ can be easily computed using the residue theorem. The denominator in Eq.~(\ref{eq:prop_mom}) can be written as:
\begin{align}
\sum_{\mu}  \tilde{p}_{\mu}^{2} + m_{q}^{2} + \frac{a^{2}}{4}\left( \sum_{\mu}\hat{p}_{\mu}^{2}\right)^{2} = -\frac{2}{a^{2}}A(\bm{p})\left( \cosh{( iap_{0})} - \cosh{(aE_{\bm{p}})}\right)~,  
\end{align}
where $A(\bm{p})$ and $E_{\bm{p}}$ are defined as
\bea
A(\bm{p}) = 1 + \frac{1}{2}a^{2}\hat{\bm{p}}^{2} ~ ,\qquad 
\cosh{(aE_{\bm{p}})} = 1+ \frac{ a^{2}B(\bm{p}) }{2A(\bm{p})} ~ ,\qquad 
B(\bm{p})  =  m_{q}^{2} + \hat{\bm{p}}^{2} + \frac{a^{2}}{2}\sum_{i<j} \hat{p}_{i}^{2}\hat{p}_{j}^{2} ~ . ~ \nonumber  
\eea
The momentum-space lattice quark propagator has two poles in the complex plane at $ip_{0} = \pm E_{\bm{p}}$, and the corresponding residue can be computed using
\begin{align}
D^{-1}(\bm{p}) \equiv \lim_{ip_{0} \to  E_{\bm{p}}}a^{2} \frac{(ip_{0} - E_{\bm{p}})}{ 2A(\bm{p})\left( \cosh{( iap_{0})} - \cosh{(aE_{\bm{p}})}\right)  } =  \frac{1}{\sqrt{B(\bm{p})\cdot\left( 4A(\bm{p}) + a^{2}B(\bm{p})\right)}}~.   
\end{align}
Using the previous results,  Eq.~(\ref{eq:prop_coord}) can be written for $t\ne 0$ as
\bea
\label{eq:corr_res}
\langle \psi_{q}(x)\bar{\psi}_{q}(y)\rangle & = & \int_{-\frac{\pi}{a}}^{\frac{\pi}{a}} \frac{d^{3}\bm{p}}{(2\pi)^{3}}e^{-E_{\bm{p}}|t|}~\frac{e^{i\bm{p}\cdot(\bm{x-y})}}{D(\bm{p})} \nonumber \\
& \cdot & \left[ \textrm{sgn}(t)\frac{\gamma_{0}}{a}\sinh{(a E_{\bm{p}})} -i\bm{\gamma}\cdot\tilde{\bm{p}} + m_{q} + ir\gamma_{5}\frac{a}{2}\left( \hat{\bm{p}}^{2} - \frac{ B(\bm{p})}{A(\bm{p})}\right)\right] ~ ,  ~  
\eea
The free-theory vector-tensor lattice  correlator
\begin{align}
C_{{\rm free}, q}(t,a) &\equiv -\frac{ie_{q}}{3}\int d^{3}{\bm x} \,\langle 0 | {\rm T}\left\{ \bar{\psi}_{q}(0)\sigma^{0j}\psi_{q}(0) \bar{\psi}_{q}(\bs{x},t)\gamma^{j}\psi_{q}(\bs{x},t)\right\} | 0 \rangle~, \nonumber \\[8pt]
&= -\frac{e_{q}}{3} \int d^{3}{\bm x} \Tr\left[ \gamma^{0}\gamma^{j} \langle \psi_{q}(0)\bar{\psi}_{q}(\bm{x},t)\rangle \gamma^{j} \langle\psi_{q}(\bm{x},t)\bar{\psi}_{q}(0)\rangle\right]~,
\end{align}
where $\Tr[.]$ is intended over both color and Dirac indices, 
can be evaluated using Eq.~(\ref{eq:corr_res}). The result is
\begin{align}
C_{{\rm free}, q}(t,a) = -\frac{8e_{q}N_{c}}{(2\pi)^{3}}\textrm{sgn}(t)\int_{-\pi/a}^{\pi/a} d^{3}\bm{p}\, e^{-2E_{\bm{p}}|t|}(am_{q})\frac{ \sinh{(a E_{\bm{p}})}}{a^{2}D^{2}{(\bm{p}})}~.
\end{align}
It is interesting to notice that the dependence on the twisted-Wilson term completely disappeared, therefore at tree level the ``tm'' and ``OS'' regularization, which as already discussed in the main text differ by the relative sign of the parameter $r$ between the forward and backward quark propagators, produce exactly the same result. 

The continuum correlator can be then obtained through the replacement
\begin{align}
\sinh( aE_{\bm p}) \to a E_{\bm p} \to a\sqrt{\bm{p}^{2}+m_{q}^{2}}, \qquad D^{2}(\bm{p}) \to 4({\bm p}^{2} + m_{q}^{2})~,
\end{align}
so that one has
\begin{align}
C_{{\rm free}, q}(t) &\equiv C_{{\rm free}, q}(t,0) = -\frac{2e_{q}N_{c}}{(2\pi)^{3}} \textrm{sgn}(t)\int_{-\infty}^{\infty} d^{3}\bm{p} \, e^{-2\sqrt{\bm{p}^{2}+ m_{q}^{2}}t} \frac{m_{q}}{\sqrt{ \bm{p}^{2} + m_{q}^{2}}}\nonumber \\[8pt]
&= -\frac{e_{q}N_{c}}{\pi^{2}} \textrm{sgn}(t)\int_{0}^{\infty} d|p| \, e^{-2\sqrt{ |p|^{2} + m_{q}^{2}}t} \frac{ m_{q} |p|^{2}}{\sqrt{ |p|^{2} + m_{q}^{2}}} \nonumber \\[8pt]
&= -\frac{e_{q}N_{c}}{2\pi^{2}}\textrm{sgn}(t)\frac{m_{q}^{2}}{t}K_{1}(2m_{q}|t|)~,\
\end{align}
where $K_{n}(x)$ are the modified Bessel functions of the second kind. From the previous equation, using $K_{1}(x) = x^{-1} +\mathcal{O}(x)$, it is immediate to show the validity of Eq.~(\ref{eq:log_divergence}).

\end{document}